\documentclass[aps,preprintnumbers,nofootinbib,floatfix]{revtex4}
\usepackage{enumitem}  
\usepackage{amssymb,bm,graphicx,wrapfig}
\usepackage{amsmath,bbold,ulem}
\usepackage{dsfont}
\usepackage{slashed}
\usepackage[dvipsnames]{xcolor}
\usepackage[utf8]{inputenc}
\usepackage{wasysym,hyperref}
\newcommand{\be}{\begin{equation}}
\newcommand{\ee}{\end{equation}}
\newcommand{\ba}{\begin{eqnarray}}
\newcommand{\ea}{\end{eqnarray}}
\newcommand{\bsub}{\begin{subequations}}
\newcommand{\esub}{\end{subequations}}
\newcommand{\la}{\langle}
\newcommand{\ra}{\rangle}

\newcommand{\red}[1]{{\color{red} #1}}

\allowdisplaybreaks

\begin{document}
\title{Quasi Parton Distribution Functions in Covariant Quark Models}
\author{
F.~Aslan$^{1,3}$,
A.~Tandogan$^2$,
P.~Schweitzer$^3$}
  \affiliation{
    $^1$ Department of Physics, Chemistry and Mathematics, Alabama A\&M University, Normal,  Alabama  35762\\
    $^2$ Department of Physics, University of Connecticut, Hartford, CT 06103, U.S.A. \\
    $^3$ Department of Physics, University of Connecticut, Storrs, CT 06269-3046, U.S.A.\\
  }
\begin{abstract}
Quasi parton distribution functions (QPDFs) are defined in terms of 
QCD fields at spacelike separations evaluated in matrix elements 
of hadrons moving with velocity $v$. These objects can~be studied 
in lattice QCD. In the limit when $v$ approaches the speed of light,
QPDFs converge to~PDFs. It is insightful to study QPDFs and their
convergence in models. In this work, we first study the QPDFs 
in a broad class of quark models characterized by one common
feature, namely the absence of gauge degrees of freedom. 
We provide general proofs for the convergence and sum rules of the 
unpolarized quark and antiquark QPDFs for both choices 
$\gamma^0$ and $\gamma^3$. 
We choose the Covariant Parton Model (CPM) as an illustration,  and
derive analytical results for the small-$x_v$ behavior of QPDFs and the 
energy-momentum tensor form factor $\bar{c}^q(t)$ at zero momentum 
transfer. These results are of interest as they correspond 
to a Wandzura-Wilczek-type approximation.
\end{abstract}

\maketitle

\section{Introduction}

Parton Distribution Functions (PDFs) \cite{Collins:1981uw} play an important 
role in the description of the internal structure of hadrons in factorizable 
deep-inelastic scattering processes \cite{Collins:1989gx}. PDFs are 
frame-independent functions defined in terms of non-local QCD field operators 
with lightlike separations, for instance, in the twist-2 unpolarized quark 
case one deals with operators of the type 
$\overline{\Psi}{ }^q(0){\cal W}(0,z)\gamma^+\Psi^q(z)$ with $z^2=0$, where 
${\cal W}$ denotes the Wilson line and $\gamma^+=(\gamma^0+\gamma^3)/\sqrt{2}$.
It was not known how to compute the $x$-dependence of PDFs in Euclidean 
space time in lattice QCD until the concept of Quasi PDFs (QPDFs) was 
introduced \cite{Ji:2013dva}. 
QPDFs are frame-dependent and defined in terms of nonlocal QCD 
operators with spacelike separations, for instance, in the quark case 
$\overline{\Psi}{ }^q(0){\cal W}(0,z)\Gamma\Psi^q(z)$ with $z^2<0$ 
where $\Gamma=\gamma^0$ or~$\gamma^3$ can be chosen, which are evaluated 
in matrix elements of hadrons moving with a velocity $v$. 
In the limit when $v$ approaches the speed of light, one recovers from 
QPDFs the frame-independent PDFs \cite{Ji:2013dva,Ji:2014gla}. 
Noteworthy is a prior discussion of the QPDF concept in a model 
application \cite{Diakonov:1997vc}. 

The two operations, (i) taking the limit $v\to1$ 
(we use natural units with $c=\hbar=1$) and (ii) carrying out the renormalization
of operators, do not commute which introduces important differences between 
PDFs and QPDFs. For instance, the latter contain higher-twist corrections 
suppressed by powers of $P_z = Mv/\sqrt{1-v^2}$ and have different 
renormalization properties which are determined systematically order 
by order in perturbative QCD matching calculations, see e.g.\ 
\cite{Xiong:2013bka,Stewart:2017tvs,Izubuchi:2018srq,Ji:2020ect}.
Lattice QCD calculations of QPDFs were reported in, e.g., 
Refs.~\cite{Lin:2014zya,Alexandrou:2015rja,Chen:2016utp,Alexandrou:2016jqi,
Alexandrou:2017huk,Chen:2017mzz,Green:2017xeu,Lin:2017ani,Alexandrou:2018pbm,
Chen:2018fwa,Alexandrou:2018eet,Liu:2018uuj,Lin:2018qky,Fan:2018dxu,
Alexandrou:2019lfo,Izubuchi:2019lyk,Bhattacharya:2020cen,
Bhattacharya:2021moj,Gao:2021dbh,Delmar:2023agv}.
For other methods to study partonic properties of hadrons 
on Euclidean lattices see
Ref.~\cite{Braun:1994jq,Detmold:2005gg,Braun:2007wv,Ma:2014jla,
Radyushkin:2016hsy,Radyushkin:2017cyf,Radyushkin:2017lvu,
Chambers:2017dov,Hansen:2017mnd,Orginos:2017kos,Ma:2017pxb}
and the reviews 
\cite{Lin:2017snn,Monahan:2018euv,Cichy:2018mum,Zhao:2018fyu,
Constantinou:2020pek,Constantinou:2022yye}.

In this work we will present a study of QPDFs in quark models.
Model studies of QPDFs have been carried out in a variety of frameworks 
and are of interest for their own sake
\cite{Gamberg:2014zwa,Bacchetta:2016zjm,Broniowski:2017wbr, 
Hobbs:2017xtq,Broniowski:2017gfp,Xu:2018eii,Bhattacharya:2018zxi,Braun:2018brg,
Bhattacharya:2019cme,Son:2019ghf,Ma:2019agv,Kock:2020frx,Son:2022qro,
Shastry:2022obb,Tan:2022kgj}. 
While lattice QCD is a reliable first-principle nonperturbative approach
to hadron structure studies, it also comes with non-trivial limitations. 
For instance, when taking the limit of QPDFs for $v\to 1$ or equivalently 
$P_z\to\infty$, power-corrections appear which are generically of the type 
$M^2/(x_v^2(1-x_v)P_z^2)$ 
\cite{Lin:2017snn,Monahan:2018euv,Cichy:2018mum,Zhao:2018fyu,
Constantinou:2020pek,Constantinou:2022yye}.
The power corrections limit the applicability of lattice calculations 
to the region of intermediate $x_v$. In models, some of the power corrections 
might be absent or exactly calculable owing to the simpler model dynamics 
and depending on the model approach. This may allow one, for instance, to 
draw conclusions about how QPDFs converge toward PDFs when the limit $v\to1$ 
is taken or make other interesting observations. If a model shares some 
features of QCD, such observations can be helpful, e.g., to better 
understand lattice studies. 

This work consists of two parts. The first part is devoted to a
general discussion of QPDFs in quark models, i.e.\ approaches with 
quark and antiquark but without gauge field degrees of freedom. 
In this part, we define the unpolarized QPDFs in quark models in 
terms of the Lorentz-invariant amplitudes $A_i^q$ and $A_i^{\bar q}$ 
which characterize the Lorentz structure of quark and antiquark correlators 
and are functions of $P\cdot k$ and $k^2$ where $P^\mu$ and $k^\mu$ denote
respectively the nucleon and quark momenta. We show how QPDFs converge 
to PDFs, and give independent derivations of sum rules for QPDFs.  
We also discuss the interesting connection 
between QPDFs and transverse momentum dependent PDFs (TMDs). We will see 
that it is convenient to formulate QPDFs as functions of hadron velocity 
$v$ rather than hadron momentum $P_z$ although both formulations are
of course equivalent. This part applies to any quark model 
framework which satisfies two very general conditions,
namely (i) Lorentz symmetry, and (ii) the amplitudes $A_i^q(P\cdot k,k^2)$ 
are either finite or, if the model exhibits any type of divergences, are 
renormalized or regularized such that Lorentz invariance is preserved.

In the second part, we will choose for definiteness a specific
model for hadron structure, namely the covariant parton model
(CPM). This model is a systematic extension of Feynman's parton 
model concept \cite{Feynman:1969ej,Feynman:1973xc} to the 
description of the partonic structure of hadrons and has been
applied with success to the description of structure functions, 
PDFs and TMDs \cite{
Zavada:1996kp,Zavada:2001bq,Zavada:2002uz,Efremov:2004tz,
Zavada:2007ww,Efremov:2009ze,DAlesio:2009cps,Zavada:2009ska,
Efremov:2009vb,Efremov:2010mt,Zavada:2011cv,Zavada:2013ola,
Zavada:2015gaa,Bastami:2020rxn,Aslan:2022wqc,Aslan:2022kmd}.
A similar approach was pursued in \cite{Blumlein:1996tp}.
Distinguished by its clarity, this model offers a convenient theoretical 
laboratory to investigate in a lucid way the properties of QPDFs.
We will derive the CPM results for QPDFs for quarks and antiquarks,
demonstrate their convergence, prove their sum rules, and derive
their small- and large-$x_v$ properties which can be studied 
analytically in that model. We will show numerical results for the 
QPDFs for the flavors $u$, $d$, $\bar u$, $\bar d$, $s$, $\bar s$
in the CPM and compare their convergence for the cases 
$\Gamma = \gamma^0$ and $\gamma^3$ in detail. The CPM results refer 
to a high renormalization scale $\mu^2$ where the partonic picture 
is justified. Remarkably, the CPM results practically constitute 
a Wandzura-Wilczek approximation for the QPDFs and automatically
entail the complete target-mass corrections.
The Appendices contain derivations of the CPM expressions for 
the electromagnetic and energy-momentum tensor form factors at 
$t=0$ required for the sum rules.

\section{QPDF\lowercase{s} in general quark models}
\label{Sec-02:general}

In this section, we discuss the definition and representation 
of QPDFs and the proofs of their properties in general quark 
model frameworks defined as effective approaches without 
explicit gauge field degrees of freedom. 
The only underlying assumptions are (i)
that the models respect Lorentz symmetry, and (ii) that, 
if divergences occur, then they are assumed to be regularized 
or renormalized such that relativistic invariance of 
the model is manifestly preserved. 
It is understood that the model results refer to 
a scale fixed in a specific way in a given model
which corresponds to the renormalization scale in QCD.
The scale dependence will not be indicated explicitly in the following.

\subsection{Notation and choice of variables}
\label{Sec:notation-Pz-vs-v}

In this work, we will denote the unpolarized QPDF as $D^a(x_v,\Gamma,v)$ 
with the two choices $\Gamma = \gamma^0$ or $\gamma^3$.
In the limit when the nucleon velocity $v\to 1$, both choices converge to
the unpolarized PDF $f_1^a(x)$. Some comments are in order.

First, it is important to distinguish the Lorentz-invariant variable $x$ of PDFs 
from the frame-dependent variable $x_v$ of QPDFs where the index $v$ stands for 
the nucleon velocity and indicates the frame dependence. If $P^\mu$ denotes the 
nucleon four-momentum and $k^\mu$ the parton four-momentum, 
then $x$ and $x_v$ are respectively defined as
\be
    x_v = \frac{k^3}{P^3} \, , \quad x = \frac{k^+}{P^+}\,
\ee
with the lightcone coordinates $k^\pm=(k^0\pm k^3)/\sqrt{2}$ and $\vec{k}_\perp=(k^1,k^2)$. 
The lightcone is selected by the hard momentum flow in a process, e.g.\ the virtual photon 
momentum in deep-inelastic electron-nucleon scattering. We will use the notations 
$k^\mu=(k^0,\,k^1,\,k^2,\,k^3)=(k^0,\vec{k}_\perp,k^3)=(k^+,\vec{k}_\perp,k^-)$
interchangeably depending on what is more convenient. Notice that $k^\mu$ is a dummy 
integration variable which will be introduced below in Sec.~\ref{Sec:introducing-A-amplitudes}
when expressing PDFs and QPDFs in terms of the quark correlator, i.e.\ no reference 
to the parton model is required when speaking of the ``parton four-momentum'' $k^\mu$.
By definition, the nucleon momentum $P^\mu$ has no transverse components.

Second, the variable $x$ is Lorentz-invariant in the sense of being boost-invariant
along the lightcone, i.e.\ we obtain the same, frame-independent PDF independently
of whether the nucleon is at rest, moves with a velocity $0<v<1$ or is boosted to 
the infinite-momentum frame $v\to 1$. The situation is different for the manifestly 
frame-dependent variable $x_v$ which can be related to $x$ as follows (see remark below)
\be\label{Eq:x-vs-xv}
    x_v = \frac{x}{v} - \frac{(1-v)}{v}\,\frac{P\cdot k}{M^2}\,,
%%    \quad x_v = \frac{k^3}{P^3}\,,
%%    \quad x   = \frac{k^+}{P^+}\,,
\ee
where $M$ is the nucleon mass. The variable $P\cdot k$ appears naturally in the 
momentum-space description of the quark correlator and is of course a Lorentz scalar. 
The Eq.~(\ref{Eq:x-vs-xv}) shows that $x_v$ is velocity-dependent and becomes equal 
to $x$ only for $v\to 1$. 
% Note that $P\cdot k$ is independent of $\vec{k}_\perp$, 
% i.e.\ the same variable $x_v$ applies to transverse momentum dependent QPDFs.
We remark that Eq.~(\ref{Eq:x-vs-xv}) is not a ``stand-alone relation'' but valid 
only under the integration over $d^4k$ with the understanding that
$x_v = k^3/P^3$ and $x=k^+/P^+$ are ``abbreviations''. The literal meaning of $x_v$ 
or $x$ (but not both at the same time) exists only in the definitions of QPDFs and 
PDFs where the quark correlator is integrated over $d^4k$ in conjunction with the
appropriate $\delta$-functions $\delta(x_v-k^3/P^3)$ or $\delta(x-k^+/P^+)$,
respectively.

Third, in QCD studies $P_z=P^3=Mv/\sqrt{1-v^2}$ is customarily used as argument
for QPDFs instead of $v$. The two variables are equivalent, but for our purposes $v$ 
is more convenient for two reasons. (i) The model expressions are more concise and elegant
with $v$. (ii) Since $v = P_z / P_0$ with $P_0=\sqrt{M^2+P_z^2}$, the variable $v$ 
corresponds effectively to the resummation of an infinite series in powers 
of $M^2/P_z^2$ (assuming $P_z > M$) which is evident from 
$v = \frac{1}{\sqrt{1+M^2/P_z^2}}$.
%      = 1 - \frac{M^2}{2P_z^2}+ \frac{3M^4}{8P_z^4} +\dots\;\,.

\newpage
\subsection{Definition and general properties}

In quark models, the unpolarized PDF and QPDF are defined as
\begin{align} \label{Eq:def-PDF-I}
    f_1^q(x) = 
    \int\frac{dz^-}{4\pi}\,e^{-ixP^+z^-}\la N_v|
    \bar{\Psi}_q(0)\gamma^+\Psi_q(z)|N_v\ra
    & \Bigl|_{z^+=0,\,\vec{z}_\perp=0\;,} 
    \\
    \label{Eq:def-quasi-I}
    D^q(x_v,\Gamma,v) = \;
    \int\frac{dz^3}{4\pi}\;e^{-ix_v^{ }P^3 z^3}\;\la N_v|
    \bar{\Psi}_q(0)\:\Gamma\:\Psi_q(z)|N_v\ra
    & \Bigl|_{z^\mu=(0,0,0,z^3)\;,}
\end{align}
where $|N_v\ra$ denotes the state of a nucleon moving with the velocity $v$ 
and momentum $P^\mu=M\,(1,0,0,v) /\sqrt{1-v^2}$ along $z$-axis.
We will often use the notation $P^\mu=(P^0,0,0,P^3)=(P_0,0,0,P_z)$.
The definitions of the corresponding antiquark distributions
follow from the relations 
\ba
    f_1^{\bar q}(x) &=& -\,f_1^q(-x)\,, \nonumber\\
    D^{\bar q}(x_v,\Gamma,v) &=& -\,D^q(-x_v,\Gamma,v)\,.
    \label{Eq:def-PDF-qbar}
\ea
PDFs are non-zero only in the region $-1 < x < 1$, while the 
QPDFs extend over the whole region $-\infty<x_v<\infty$. 

The PDF in Eq.~(\ref{Eq:def-PDF-I}) is invariant under 
boosts along $z$-axis, i.e.\ independent of the nucleon 
velocity $v$. This allows one 
to evaluate PDFs in any frame including nucleon rest frame,
but at the price of  light-like separated quark fields
which hinders an evaluation in Euclidean space-time 
in lattice QCD.  In contrast to this, in the QPDF in 
Eq.~(\ref{Eq:def-quasi-I}), the separation of the quark fields  
is space-like which makes a computation on a Euclidean lattice 
possible, but the QPDF is frame dependent and the 
connection to a PDF requires the limit $v\to 1$.

Except for the Wilson line absent in quark models,
Eq.~(\ref{Eq:def-quasi-I}) corresponds to the QCD definition 
of Ref.~\cite{Ji:2013dva} where QPDFs
were introduced with applications to lattice QCD in mind.
QPDFs were introduced prior to that in Ref.~\cite{Diakonov:1997vc} 
for a different purpose, namely to prove that, in the chiral 
quark-soliton model, the PDFs obtained from evaluating the
(i) operator definition in Eq.~(\ref{Eq:def-PDF-I}) are equivalent to 
(ii) probabilistic parton densities in infinite-momentum frame.
To the best of our knowledge, this is the only model where this was
explicitly proven~\cite{Diakonov:1997vc}.
 
The QPDFs have the following important properties.
In the limit of the nucleon velocity $v$ approaching the speed 
of light, the QPDFs and PDFs are connected as follows
\be\label{Eq:general-limit-quasi-PDF}
      \lim\limits_{v\to 1} x_v = x 
      \, , \quad \lim\limits_{v\to 1} D^q(x_v,\Gamma,v) = f_1^q(x)
      \, , \quad \Gamma = \gamma^0,\;\gamma^3\,. 
\ee
Other important properties are the sum rules for the first and second 
Mellin moments \cite{Bhattacharya:2019cme}. For PDFs we have
\ba
      % \int_0^1dx\,\biggl(f_1^q(x)-f_1^{\bar q}(x)\biggr) &\equiv& 
      \int_{-1}^1 dx\,f_1^q(x)      
      &=&  N^q \;,
      \label{Eq:general-mom1-PDF}\\     
      % \int_0^1dx\,x\biggl(f_1^q(x)+f_1^{\bar q}(x)\biggr) &\equiv& 
      \int_{-1}^1 dx\,x\,f_1^q(x) 
      &=&  A^q(0) \;, \label{Eq:general-mom2-PDF}
\ea
where $N^q$ denotes the number of valence quarks of flavor $q$, and 
$A^q(t)$ is a form factor of the energy-momentum tensor with the 
property that summation over all quark flavors and gluons yields
$\sum_a A^a(0)=1$ \cite{Polyakov:2018zvc,Burkert:2023wzr}. 
For the QPDFs, the corresponding sum rules in the case $\Gamma=\gamma^0$
are given by  
\begin{align}
    %  \int_0^\infty dx_v\,\biggl(D^q(x_v,\gamma^0,v)-D^{\bar q}(x_v,\gamma^0,v) \biggr)
    \int_{-\infty}^\infty dx_v\,D^q(x_v,\gamma^0,v) 
    & = \frac{N^q}{v}\,,
      \label{Eq:general-mom1-quasi-0}\\
    %  \int_0^\infty dx_v\,x\biggl(D^q(x_v,\gamma^0,v)+D^{\bar q}(x_v,\gamma^0,v) \biggr)
    \int_{-\infty}^\infty dx_v\,x_v\,D^q(x_v,\gamma^0,v)
    & =  \frac{A^q(0)}{v}\,, \hspace{6mm}
    \label{Eq:general-mom2-quasi-0}
\end{align}
while in the case $\Gamma=\gamma^3$ they are given by  
\begin{align}
    %  \int_0^\infty dx_v\,\biggl(D^q(x_v,\gamma^3,v)-D^{\bar q}(x_v,\gamma^3,v) \biggr)
    \int_{-\infty}^\infty dx_v\,D^q(x_v,\gamma^3,v)
    & = \,N^q \;,
    \label{Eq:general-mom1-quasi-3}\\   
    %  \int_0^\infty dx_v\,x\biggl(D^q(x_v,\gamma^3,v)+D^{\bar q}(x_v,\gamma^3,v) \biggr)
    \int_{-\infty}^\infty dx_v\,x_v\,D^q(x_v,\gamma^3,v) 
    & =  A^q(0) - \frac{1-v^2}{v^2}\,{\bar c}^q(0)\;,
    \label{Eq:general-mom2-quasi-3}
\end{align}
where $\bar{c}^a(t)$ is the form factor expressing the non-conservation 
of the individual quark and gluon contributions to the energy-momentum
tensor which vanishes upon the summation over all constituents 
$\sum_a \bar{c}^a(t)=0$ \cite{Polyakov:2018zvc,Burkert:2023wzr}.

% Notice that in a given model, the PDFs and QPDFs may contain 
% UV divergences which we assume to be either regularized or renormalized
% depending on the framework. 
% In non-renormalizable, effective, low energy approaches,
% UV divergences are removed by subtracting off momenta above a 
% certain UV cutoff within an appropriate regularization scheme
% in which case the UV cutoff sets the scale to which the model 
% results refer, see e.g.\ the model of 
% Refs.~\cite{Diakonov:1997vc,Son:2019ghf}.
% In renormalizable approaches, the removal of UV divergences 
% through a regularization and renormalization procedure introduces 
% a renormalization scale dependence. In either case, the 
% theoretical results refer to a scale which we shall not indicate 
% for notational brevity throughout this work, but when needed we will comment on the scale.

In Refs.~\cite{Radyushkin:2016hsy,Radyushkin:2017cyf} a relation between one of the 
QPDFs and the TMD $f_1^q(x,\vec{k}{ }_T^2)$ 
was derived which, in our convention, is given by the formula (see App.~\ref{App:conventions} 
for conventions)
\be\label{Eq:Radyushkin-formula}
    D^q(x_v,\gamma^0,v) 
    = P_0\,\int_{-1}^1dx \int_{-\infty}^\infty
    dk_1\,f_1^q\left(x,\,k_1^2+(x_v-x)^2P_z^2\right)\,.
\ee
The TMD entering this formula is defined in QCD with a straight gauge link
and constitutes a field-theoretical object which can be computed in lattice QCD
but differs from what is measured in deep-inelastic scattering reactions. 
In quark models where gauge links are not present, the "usual" 
TMD enters Eq.~(\ref{Eq:Radyushkin-formula}).

\subsection{\boldmath Expressions in terms of Lorentz-invariant 
amplitudes $A_i^q(P\cdot k,k^2)$}
\label{Sec:introducing-A-amplitudes}

For our purposes, it will be convenient to rewrite the 
definitions in Eqs.~(\ref{Eq:def-PDF-I},~\ref{Eq:def-quasi-I})
in terms of the unintegrated quark correlator which is defined 
in quark models as follows \cite{Mulders:1995dh}
\be\label{Eq:correlator-q}
    \Phi_{ij}^q(k,P,S) =
    \int \frac{\mathrm{d}^4z}{(2\pi)^4}\;\mathrm{e}^{i k\cdot z}\,
    \langle N_v|\,\overline{\Psi}_j^{\,q}(0)\;\Psi_i^q(z)\,
    |N_v\rangle\,. 
\ee
There is an analog definition for the antiquark correlator 
which is related to the quark correlator by the field-theoretic
relation  $\Phi_{ij}^{\bar q}(k,P,S) = - \,\Phi_{ij}^q(-k,P,S)$
\cite{Mulders:1995dh}.
In quark models the correlator is a function of the quark momentum
$k^\mu$, the nucleon momentum $P^\mu$ and nucleon polarization
vector $S^\mu$. In QCD the correlator definitions include Wilson~lines. 
As a consequence the correlators in QCD depend on an additional 
four-vector $n^\mu$ which specifies a direction along which the Wilson 
lines are chosen according to the factorization theorem of a specific 
process \cite{Goeke:2005hb}.
By making use of the correlator in Eq.~(\ref{Eq:correlator-q}) the 
PDF and QPDF in Eqs.~(\ref{Eq:def-PDF-I},~\ref{Eq:def-quasi-I})
are given by
\begin{align} \label{Eq:def-PDF-II}
    f_1^q(x) = \frac{1}{2P^+}\int d^4k\;{\rm tr}
    & \bigl[\Phi^q(k,P,S)\gamma^+\bigr]\,
    \delta\bigl(x-\frac{k^+}{P^+}\bigr) \, ,
    \\
    \label{Eq:def-quasi-II}
    D^q(x_v,\Gamma,v) = \frac{1}{2P^3}\int d^4k\;{\rm tr}
    & \bigl[\Phi^q(k,P,S)\;\Gamma\;\bigr]\,
    \delta\bigl(x_v-\frac{k^3}{P^3}\bigr) \, .
\end{align}
    In QCD the fully unintegrated quark correlator with a general
	gauge-link geometry can be decomposed into 32 linearly independent
	Lorentz structures constructed from the four-vectors
	$k^\mu$, $P^\mu$, $S^\mu$, and $n^\mu$, where $n^\mu$ specifies an
	independent gauge-link direction \cite{Goeke:2005hb}. 
    In quark models without explicit gauge-field degrees of freedom, the
	Wilson-line-dependent structures are absent, and the correlator is
	described by only 12 independent amplitudes,
	$A_i^q=A_i^q(P\cdot k,k^2)$, which are functions of the Lorentz scalars
	$P\cdot k$ and $k^2$ (which we often do not indicate explicitly for
	notational simplicity) \cite{Mulders:1995dh}. The reduced number of
	linearly independent Lorentz structures in quark models implies
	quark-model relations among leading- and subleading-twist
	transverse-momentum-dependent PDFs; see, for example,
	Refs.~\cite{Metz:2008ib,Teckentrup:2009tk}. 
The decomposition of the quark correlator in quark models
(analog for the antiquark correlator) is as follows
\cite{Boer:1997nt} 
\begin{align}
   \label{Eq:correlator-decompose}
   \Phi^a(k,P,S) \; 
   & =
     MA_1^a + \slashed{P} A_2^a + \slashed{k}A_3^a 
     + \frac{i}{2M} \; [\slashed{P},\slashed{k}] \; A_4^a
     + i (k\cdot S) \gamma_5 \; A_5^a 
     + M\slashed{S} \gamma_5 \; A_6^a
     + \frac{(k\cdot S)}{M} \slashed{P} \gamma_5 \; A_7^a
     + \frac{(k\cdot S)}{M} \slashed{k} \gamma_5 \; A_8^a
 \nonumber\\
   & + \frac{[\slashed{P},\slashed{S}]}{2}\gamma_5\; A_9^a 
     + \frac{[\slashed{k},\slashed{S}]}{2} \gamma_5 \; A_{10}^a 
     + \frac{(k\cdot S)}{2M^2} [\slashed{P},\slashed{k}] \gamma_5 \; A_{11}^a
    + \frac{1}{M} \varepsilon^{\mu\nu\rho\sigma} \gamma_{\mu} P_{\nu} k_{\rho} S_{\sigma} \; A_{12}^a \,.
\end{align}
Eq.~(\ref{Eq:correlator-decompose}) 
applies generally to {\it all} quark models without explicit gauge-field degrees of freedom.
Note that in QCD and models with gluons, besides the 
12 $A_i^q(P\cdot k,k^2)$ amplitudes in Eq.~(\ref{Eq:correlator-decompose}), 
20 additional $B_i^q(P\cdot k,k^2)$ amplitudes appear due to the
vector $n^\mu$ \cite{Goeke:2005hb}.

In our case, we are interested in the projection of the correlator
on the Dirac structure $\gamma^\mu$ which yields
\be\label{Eq:tr-phi-gamma-mu}
    {\rm tr}\bigl[\Phi^q(k,P,S)\,\gamma^\mu\bigr] =
    4\,\biggl(P^\mu\,A_2^q+ k^\mu\,A_3^q\biggr)\,.
\ee
In this way, the expressions 
for the unpolarized PDF and QPDF are given by
\begin{align} \label{Eq:def-PDF-III}
    f_1^q(x) = 2\int d^4k\;
    & \biggl(A_2^q+x\,A_3^q\biggr)\,
    \delta\bigl(x-\frac{k^+}{P^+}\bigr) \, ,
    \\
    \label{Eq:def-quasi-III}
    D^q(x_v,\gamma^\mu,v) = 2\int d^4k\;
    & \biggl(\frac{P^\mu}{P^3}\, A_2^q
            +\frac{k^\mu}{P^3} \,A_3^q\biggr)\,
    \delta\bigl(x_v-\frac{k^3}{P^3}\bigr) \, ,
    \quad \mu = 0,\:3\,.
\end{align}
The corresponding expressions for antiquarks follow
from Eq.~(\ref{Eq:def-PDF-qbar}) by exploring
$\Phi_{ij}^{\bar q}(k,P,S) = - \,\Phi_{ij}^q(-k,P,S)$
\cite{Mulders:1995dh} which implies for the amplitudes
$A_2^q(-P\cdot k,k^2)=-A_2^{\bar q}(P\cdot k,k^2)$ and
$A_3^q(-P\cdot k,k^2)= A_3^{\bar q}(P\cdot k,k^2)$
\cite{Aslan:2022wqc}.

\subsection{Infinite momentum frame limit}

In this section we show that the quark model expressions 
for QPDFs (\ref{Eq:def-quasi-III}) yield the PDF 
(\ref{Eq:def-PDF-III}) in the limit~$v\to1$.
The amplitudes $A_i^q$, their arguments $P\cdot k$ and $k^2$,
and $d^4k$ are Lorentz scalars which can be evaluated in any frame. 
But for the specific components of $P^\mu$ and $k^\mu$ we must choose
a specific frame. 
It is convenient to choose a frame in which the nucleon
and quark momenta are the results of a boost from the 
nucleon rest frame where we have 
\be\label{Eq:P-k-rest-frame}
    P^\mu=(M,0,0,0)\,, \quad
    k^\mu=(k^0,k^1,k^2,k^3) \qquad \mbox{(rest frame)},
\ee
to a frame where the nucleon moves with velocity $v$.
The nucleon and quark four-momenta in this frame are given by 
\be
\label{Eq:P-k-of-v}
    P^\mu = \biggl(\frac{M}{\sqrt{1-v^2}},\,
    0,0, \frac{Mv}{\sqrt{1-v^2}}\biggr)\,,\quad
    k^\mu = \biggl(\frac{k^0+v\,k^3}{\sqrt{1-v^2}},\,
    k^1,\,k^2, \frac{k^3+v\,k^0}{\sqrt{1-v^2}}\biggr)
    \qquad \mbox{(boost from rest frame)}.
\ee
The evaluation of the model expressions is facilitated by the fact 
that the amplitudes $A_i^q$, as well as  $P\cdot k$,
$k^2$ and $d^4k$ are Lorentz scalars, which means that the integrations can be carried out
in any frame. We will make use of this property. 
For instance, evaluating $P\cdot k$ in the frame (\ref{Eq:P-k-of-v}) 
yields $P\cdot k=Mk^0$ which is equivalent to evaluating this scalar product 
in the rest frame (\ref{Eq:P-k-rest-frame}) although the nucleon of 
course moves with the velocity $v$. This illustrates that 
(i) $P\cdot k$ can of course be evaluated in any frame including the nucleon rest frame, 
(ii) as a Lorentz scalar it must be velocity-independent as is the case. 
In the frame (\ref{Eq:P-k-of-v}) the arguments of amplitudes are given by 
$A_i^q=A_i^q(Mk^0,k^2)$ which has a practical advantage: in this
frame the amplitudes are rotationally symmetric and even functions of the 
3-momentum components $k^i$. 
Despite choosing a preferred reference frame for the $d^4k$-integration, 
the PDF expressions are nevertheless frame independent, 
and those for QPDFs have the correct frame (velocity) dependence.

From Eq.~(\ref{Eq:P-k-of-v}) we see 
that $k^+/P^+ = (k^0+k^3)/M $ is velocity-independent 
and hence so is the PDF which is expected. By exploring 
the $\delta$-function in Eq.~(\ref{Eq:def-PDF-III}) to 
replace $x$ by $k^+/P^+=(k^0+k^3)/M$ we obtain for the PDF
\be\label{Eq:PDF-reference-expression}
    f_1^q(x) = 2\int d^4k\;
    \biggl(A_2^q+\frac{k^0+k^3}{M}\,A_3^q\biggr)\,
    \delta\bigl(x-\frac{k^0+k^3}{M}\bigr) \, ,
\ee
which will be a helpful reference expression in the following.
For the QPDFs with $\Gamma = \gamma^0,\,\gamma^3$ we obtain 
\begin{align} 
    \label{Eq:def-quasi-0-v-limit}
    D^q(x_v,\gamma^0,v) = 2\int d^4k\;
    & \biggl(\frac{A_2^q}{v}
          +\frac{k^0+v\,k^3}{vM} \,A_3^q\biggr)\,
    \delta\bigl(x_v-\frac{k^3+vk^0}{vM}\bigr) \, ,\\
    \label{Eq:def-quasi-3-v-limit}
    D^q(x_v,\gamma^3,v) = 2\int d^4k\;
    & \biggl(A_2^q
          +\frac{k^3+v\,k^0}{vM} \,A_3^q\biggr)\,
    \delta\bigl(x_v-\frac{k^3+vk^0}{vM}\bigr) \, .
\end{align}
In both expressions it is straightforward to take  
the limit $v\to1$ and we recover in both cases the PDF 
in Eq.~(\ref{Eq:PDF-reference-expression}).

%%%%%%%%%%%%%%%%%%%%%%%%%%%%%%%%%%%%%%%%%%%%%%%%
\subsection{Flavor number sum rule}
%%%%%%%%%%%%%%%%%%%%%%%%%%%%%%%%%%%%%%%%%%%%%%%%

In this section, we show how the flavor number sum rule is satisfied
in quark models. Let us first derive the sum rule for the PDF. 
In our reference frame (\ref{Eq:P-k-of-v}) in the representation of 
Eq.~(\ref{Eq:PDF-reference-expression}) we obtain 
\be\label{Eq:mom1-PDF}
    \int dx\;f_1^q(x) =  2\int d^4k\;
    \biggl(A_2^q+\frac{k^0}{M}\,A_3^q\biggr) \,,
\ee
where a term proportional to $k^3\,A_3^q$ dropped out due to 
rotational symmetry of the integrand which emerges after 
$x$-integration when the $z$-axis is no longer singled out.
Alternatively, after $x$-integration one may substitute under 
the $d^4k$-integral $k^3\to(-k^3)$ to see that this term changes 
sign and drops out. Keep in mind that in the frame (\ref{Eq:P-k-of-v})
we have $A_i^q=A_i^q(Mk^0,k^2)$ which are even functions of spatial
components $k^i$ and invariant under rotations. The result can be 
formulated also in a covariant way, see App.~\ref{App:em-FF}, where 
we will also show how the expression on the right-hand-side of
Eq.~(\ref{Eq:mom1-PDF}) is related to the flavor quantum number 
$N^q$ in Eq.~(\ref{Eq:mom1-PDF}).

We now derive the flavor sum rule for the case $\Gamma = \gamma^0$.
Using the frame (\ref{Eq:P-k-of-v}) and exploring the representation 
in Eq.~(\ref{Eq:def-quasi-0-v-limit}) and integrating over $x$ we find
\be\label{Eq:mom1-quasi-0}
    \int dx_v\,D^q(x_v,\gamma^0,v) 
    = \frac{2}{v}\int d^4k\; \biggl(A_2^q
    + \frac{k^0}{M} \,A_3^q\biggr)
    = \frac1v\,\int dx\;f_1^q(x)
\ee
where again for symmetry reasons a term proportional to $k^3$
dropped out. The remaining integral coincides with that in 
Eq.~(\ref{Eq:mom1-PDF}) up to a prefactor of $1/v$ which is
the expected velocity dependence for this sum rule, see
Eq.~(\ref{Eq:general-mom1-quasi-0}).

We proceed analogously for $\Gamma=\gamma^3$. In the frame 
(\ref{Eq:P-k-of-v}), we obtain after the $x$-integration of 
Eq.~(\ref{Eq:def-quasi-3-v-limit}) the result
\be\label{Eq:mom1-quasi-3}
    \int dx_v D^q(x_v,\gamma^3,v) = 
    2\int d^4k\; \biggl(A_2^q+\frac{k^0}{M} \,A_3^q\biggr)
    = \int dx\;f_1^q(x)\,,
\ee
where again a term proportional to $k^3$ dropped out, and
the resulting integral agrees with Eq.~(\ref{Eq:mom1-PDF}) 
which is velocity independent as expected for this sum rule,
see Eq.~(\ref{Eq:general-mom1-quasi-3}). 

The above derivations show that, if in a quark model the PDF flavor sum
rule $\int_{-1}^1 dx\,f_1^q(x)=N^q$ is fulfilled, then so are the sum 
rules (\ref{Eq:general-mom1-quasi-0},~\ref{Eq:general-mom1-quasi-3})
for the QPDFs. It remains to be proven that the PDF complies with 
the quark flavor sum rule. This is done in App.~\ref{App:em-FF} where 
by evaluating the Dirac form factor $F_1^q(t)$ we show that  
$2\int d^4k\;(A_2^q+\frac{k^0}{M}\,A_3^q)=N^q$ which coincides 
with the above-encountered integral expressions and completes 
the proof.

%%%%%%%%%%%%%%%%%%%%%%%%%%%%%%%%%%%%%%%%%%%%%%%%%%
\subsection{Momentum sum rule}
\label{Sec:mom-sum-rule}
%%%%%%%%%%%%%%%%%%%%%%%%%%%%%%%%%%%%%%%%%%%%%%%%%%

Next we investigate the second Mellin moment sum rule. 
For PDFs it corresponds to the momentum sum rule. 
For QPDFs there is no direct correspondence to momentum carried by 
partons, but we shall nevertheless loosely refer to this sum rule by 
the same name. In the PDF case, we obtain from 
Eq.~(\ref{Eq:def-quasi-0-v-limit}) the result
\ba\label{Eq:mom2-PDF}
    \int dx\;x\,f_1^q(x) 
    = 2\int d^4k\;\biggl(\frac{k^0}{M}A_2^q
    + \frac{k_0^2+\frac13\,\vec{k}^{\,2}}{M^2}\,A_3^q\biggr)  \,,
\ea
where terms  proportional to $k^3 A_2^q$ and $k^0k^3 A_3^q$
dropped out for symmetry reasons.
Moreover we use the fact that under the spatial part of the 
$d^4k$ integral one can replace $(k^3)^2\to\frac13\,\vec{k}^{\,2}$
for symmetry reasons. 

Having derived the result for the PDF, we now consider
the QPDF in the case $\Gamma = \gamma^0$ in 
Eq.~(\ref{Eq:def-quasi-0-v-limit}). After the 
$x$-integration terms proportional to $k^3 A_2^q$ 
and $k^0k^3 A_3^q$ drop out also in this case, and we obtain 
\be \label{Eq:mom2-quasi-0}
    \int dx_v\;x_v\,D^q(x_v,\gamma^0,v) = \frac{2}{v}\int d^4k\;
    \biggl(\frac{k^0}{M}\,A_2^q
    +\frac{k_0^2+\frac13\,\vec{k}^{\,2}}{M^2}\,A_3^q\biggr)\,,
\ee
which corresponds to the result in Eq.~(\ref{Eq:general-mom2-quasi-0})
including the expected velocity dependence. After very similar steps, 
we obtain for the case $\Gamma = \gamma^3$ the result
\be \label{Eq:mom2-quasi-3}
    \int dx_v\;x_v\,D^q(x_v,\gamma^3,v) 
    = 2\int d^4k\;\biggl(\frac{k^0}{M}\,A_2^q
    + \frac{k_0^2+\frac13\,\vec{k}^{\,2}}{M^2}\,A_3^q\biggr)
    - \frac{1-v^2}{v^2}\;
    \int d^4k\;\biggl(-\frac23\;\frac{\vec{k}^{\,2}}{M^2}\biggr)\,A_3^q 
    \, .
\ee
We recover the expected velocity dependence of the sum rule for 
$\Gamma=\gamma^3$ in agreement with Eq.~(\ref{Eq:general-mom2-quasi-3}). 

In order to complete the proof, it is necessary to show that the
above integral expressions correspond to the respective form factors
of the energy-momentum tensor at zero-momentum transfer. This step is 
shown  in App.~\ref{App:EMT-FF} which completes the proof of the sum 
rules (\ref{Eq:general-mom2-PDF}, \ref{Eq:general-mom2-quasi-0}, 
\ref{Eq:general-mom2-quasi-3}) in quark models.

\subsection{Proof of the Radyushkin formula}
\label{Sec:Radyushkin-formula}

It is instructive to review the proof of Eq.~(\ref{Eq:Radyushkin-formula})
as formulated in Ref.~\cite{Radyushkin:2017cyf} showing that it is a
consequence of Lorentz invariance (the earlier proof
\cite{Radyushkin:2016hsy} is based on a Nakanishi-type representation).
The starting point is the correlator (\ref{Eq:correlator-q}) projected on
$\gamma^\mu$ in coordinate space representation. Let this correlator 
$C^\mu(P,z)$ be denoted as
\be\label{Eq:correlator-coordinate-space}
   C^\mu(P,z) = \langle N|\,\overline{\Psi}^{\,q}(0)\,\gamma^\mu\,\Psi^q(z)\,|N\rangle\,
   = 2 P^\mu {\cal M}^q(\nu,z^2) + 2 z^\mu M^2{\cal J}^q(\nu,z^2) \, ,
   \qquad \nu = -P\cdot z \, ,
\ee
with the decomposition dictated by Lorentz invariance: the Lorentz index of $C^\mu(P,z)$
can be carried by the 4-vectors $P^\mu$ and $z^\mu$, and the scalar amplitudes
${\cal M}^q$ and ${\cal J}^q$ can only be functions of $P\cdot z$ and $z^2$.
The variable $\nu = -P\cdot z$ is called Ioffe time.
The factor $M^2$ is introduced for convenience such that ${\cal M}^q$ and
${\cal J}^q$ have the same dimension. 

The PDF is defined by setting the index $\mu=+$ and
$z^\mu=(0,\vec{0}_\perp,z^-)$ implying $\nu=-P^+z^-$ and $z^2=0$.
The TMD is defined for $\mu=+$ and $z^\mu=(0,\vec{z}_\perp,z^-)$
which implies $\nu=-P^+z^-$ and $z^2=-\vec{z}_\perp^2$.
$D^q(x_v,\gamma^0,v)$ is defined for $z^\mu=(0,0,0,z^3)$ such that
$\nu=P^3z^3$ or $z^3 = \nu/P^3$ and $z^2=-\nu^2/P_3^2$.
The definitions are given by \cite{Radyushkin:2016hsy,Radyushkin:2017cyf}
\bsub\label{Eqs:M-I}\ba
   {\cal M}^q(\nu,0) &=& \int_{-1}^1dx\,e^{i\nu x} f_1^q(x)\, , \label{Eq:M-PDF-I}\\
   {\cal M}^q(\nu,-\vec{z}_\perp^2) &=& \int_{-1}^1dx\,e^{i\nu x} \int d^2k_\perp
   e^{i\vec{k}_\perp\cdot\vec{z}_\perp}f_1^q(x,\vec{k}_\perp^2)\, , \label{Eq:M-TMD-I}\\
   {\cal M}^q(\nu,-\tfrac{\nu^2}{P_3^2}) &=& \int_{-\infty}^\infty dx_v\,e^{i\nu x_v}
   v\,D^q(x_v,\gamma^0,v)\,. \label{Eq:M-QPDF-I}
\ea\esub
Notice that the Fourier transforms with respect to $x$ in
Eqs.~(\ref{Eq:M-PDF-I},~\ref{Eq:M-TMD-I}) have support only in
the region $-1<x<1$. Inverting the Fourier transforms in
Eqs.~(\ref{Eqs:M-I}) yields 
\bsub\label{Eqs:M-II}\ba
   f_1^q(x)\, &=& \int_{-\infty}^\infty\frac{d\nu}{2\pi}\,e^{-i\nu x} {\cal M}^q(\nu,0),
   \label{Eq:M-PDF-II}\\ 
   f_1^q(x,\vec{k}_\perp^2) &=& \int_{-\infty}^\infty\frac{d\nu}{2\pi}\,e^{-i\nu x}\int
   \frac{d^2z_\perp}{(2\pi)^2}\, e^{-i\vec{k}_\perp\cdot\vec{z}_\perp}{\cal M}^q(\nu,-\vec{z}_\perp^2)\, ,
   \label{Eq:M-TMD-II}\\
   v\,D^q(x_v,\gamma^0,v) &=& \int_{-\infty}^\infty \frac{d\nu}{2\pi} \,e^{-i\nu x_v}
   {\cal M}^q(\nu,-\tfrac{\nu^2}{P_3^2}) \,. \label{Eq:M-QPDF-II}
\ea\esub
The Radyushkin formula (\ref{Eq:Radyushkin-formula}) can be derived as follows.
For ${\cal M}^q(\nu,-\tfrac{\nu^2}{P_3^2})$ in Eq.~(\ref{Eq:M-QPDF-II}) one chooses
the representation
\be
   {\cal M}^q(\nu,-\tfrac{\nu^2}{P_3^2})= {\cal M}^q(\nu,-\vec{z}_\perp^{\,2})\biggl|_{
     {\rm for}\;\vec{z}_\perp=(0,\tfrac{\nu}{P^3})}\,.
\ee
Inserting this representation in Eq.~(\ref{Eq:M-QPDF-II}) and carrying
out the integrations over $d\nu$ and $d^2z_\perp$ yields 
Eq.~(\ref{Eq:Radyushkin-formula}) \cite{Radyushkin:2017cyf}.

Several comments are in order. First, the amplitude ${\cal J}^q(\nu,z^2)$ is higher
twist and ``it is better to get rid of it'' \cite{Radyushkin:2017cyf}. It drops out
in PDFs and TMDs automatically by focusing in $C^\mu(\nu,z^2)$ on the index $\mu=+$
while $z^+=0$. For QPDFs, ${\cal J}^q(\nu,z^2)$ drops out only for $\Gamma=\gamma^0$
when choosing $\mu=0$ while $z^\mu=(0,0,0,z^3)$, but not in the case $\Gamma=\gamma^3$.
Second, physically $z^3 = \tfrac{\nu}{P_3}$
and $\vec{z}_\perp$ are along different spatial directions, but Eq.~(\ref{Eq:Radyushkin-formula})
is mathematically correct independently of the physical meaning of the variables. We will
check this below by an explicit model calculation. Third, the factor of $v$ in front of
$D^q(x_v,\gamma^0,v)$ is due to different conventions in this work vs
\cite{Radyushkin:2016hsy,Radyushkin:2017cyf}, see App.~\ref{App:conventions}.
Fourth, we recall that in QCD the TMD in (\ref{Eq:Radyushkin-formula}) is defined with a
straight gauge link and corresponds to a well-defined field theoretic object which can,
e.g., be computed on the lattice but does not correspond to the TMDs that enter 
deep-inelastic processes. In contrast to this in quark models, where gauge links are absent,
a regular (model) TMD enters in Eq.~(\ref{Eq:Radyushkin-formula}).

In order to prove the Radyushkin formula in our approach, we first need the expression
for $f_1^q(x,\vec{k}{ }_\perp^{\,2})$ in terms of the $A_i^q=A_i^q(P\cdot k,k^2)$
amplitudes which is given by
\be\label{Eq:def-TMD}
    f_1^q(x,\vec{k}_\perp^{\,2}) = 2\int\!\!\!\!\int dk^0dk^3\;
    \biggl(A_2^q+x\,A_3^q\biggr)\,
    \delta\bigl(x-\frac{k^+}{P^+}\bigr) \, ,
\ee
where the dependence on $\vec{k}_\perp^{\,2}$ enters through the argument 
$k^2=k_0^2-\vec{k}_\perp^{\,2}-k_3^2$ of the $A_i^q(P\cdot k,k^2)$ amplitudes. 
Inserting this expression in Eq.~(\ref{Eq:Radyushkin-formula}) yields
\ba\label{Eq:toward-Radyushkin-I}
    P_0\,\int_{-1}^1dx \int_{-\infty}^\infty
    dk_1\,f_1^q\left(x,\,k_1^2+(x_v-x)^2P_z^2\right)
    &=&
    2P_0\,\int_{-1}^1dx \int d^4k\,
    \biggl(A_2^q+x\,A_3^q\biggr)\,
    \delta\biggl(x-\frac{k^+}{P^+}\biggr)\,
    \delta\biggl(k^2-(x_v-x)P^3\biggr) \nonumber\\
    &=&
    \frac{2}{v}\,\int d^4k
    \biggl(A_2^q+\frac{k^+}{P^+}\,A_3^q\biggr)\,
    \delta\biggl(x_v-\frac{k^+}{P^+}-\frac{k^2}{P^3}\biggr) 
\ea
where we implemented the condition that the transverse component $k^2$ in the TMD was
fixed to the value $(x_v-x)P^3$ through integration over the $\delta$-function
$\delta(k^2-(x_v-x)P^3)=\frac1{P^3}\delta(x_v-\frac{k^+}{P^+}-\frac{k^2}{P^3})$
and used $v=\frac{P^3}{P^0}$. 

The expression in Eq.~(\ref{Eq:toward-Radyushkin-I}) is the QPDF
$D^q(x_v,v,\gamma^0,v)$ albeit not in the representation of Eq.~(\ref{Eq:def-quasi-III})
but in an equivalent representation. To demonstrate this, we follow some steps of 
the proof of Ref.~\cite{Radyushkin:2017cyf}. For that we express the coordinate
space correlator $C^\mu(P,z)$ in Eq.~(\ref{Eq:correlator-coordinate-space}) through our
momentum space amplitudes $A_i^q=A_i^q(P\cdot k,k^2)$. The former corresponds to the Fourier
transform of the momentum space correlator in Eq.~(\ref{Eq:tr-phi-gamma-mu}) such that
we have
\be\label{Eq:toward-Radyushkin-II}
   C^\mu(P,z) =  4\int d^4k\,e^{-ik\cdot z}\biggl(P^\mu A_2^q + k^\mu  A_3^q\biggr) \,.
\ee
Choosing for the Lorentz index $\mu=+$ and $z^\mu=(0,\vec{z}_\perp,z^-)$
the amplitude ${\cal J}^q$ drops out, see above, and we obtain for
${\cal M}^q(\nu,-\vec{z}^{\,2})$ the representation
\be\label{Eq:toward-Radyushkin-III}
   {\cal M}^q(\nu,-\vec{z}^{\,2}) = 2\int d^4k\,
   e^{i\nu \frac{k^+}{P^+}+i\vec{k}_\perp\cdot\vec{z}_\perp}\,
   \biggl(A_2^q + \frac{k^+}{P^+} A_3^q\biggr) \,.
\ee
Choosing $\vec{z}_\perp=(0,\tfrac{\nu}{P_3})$ in Eq.~(\ref{Eq:toward-Radyushkin-III})
and inserting the so obtained representation for
${\cal M}^q(\nu,-\tfrac{\nu^2}{P_3^2})$ in Eq.~(\ref{Eq:M-QPDF-II}) yields after
integrating over $d\nu$ and $d^2z_\perp$ the following equation
\be\label{Eq:toward-Radyushkin-IV}
    D^q(x_v,\gamma^0,v) = \frac{2}{v}\int d^4k\;
    \biggl(A_2^q +\frac{k^+}{P^+} \,A_3^q\biggr)\,
    \delta\biggl(x_v-\frac{k^+}{P^+}-\frac{k^2}{P^3}\biggr) \,.
\ee
Alternatively, setting $\mu=+$ and $z^\mu=(0,0,0,z^3)$ in 
(\ref{Eq:toward-Radyushkin-II}) and plugging the so-obtained representation 
of ${\cal M}^q(\nu,-z_3^2)$ with $z^3=\tfrac{\nu}{P^3}$ into 
Eq.~(\ref{Eq:M-QPDF-II}), yields the result in Eq.~(\ref{Eq:def-quasi-III})  
demonstrating that the two expressions for $D^q(x_v,\gamma^0,v)$ in
Eqs.~(\ref{Eq:def-quasi-III},~\ref{Eq:toward-Radyushkin-I}) are equivalent
which completes the proof. While it is not an independent proof but rather an 
``adaptation'' of the proof \cite{Radyushkin:2017cyf} to our momentum-space 
based approach, this shows that Eq.~(\ref{Eq:Radyushkin-formula}) must hold 
also in quark models. This presents an important cross-check for model computations. 

In Ref.~\cite{Radyushkin:2017cyf} it was concluded that 
Eq.~(\ref{Eq:Radyushkin-formula}) is based on Lorentz invariance. 
Since we describe $f_1^q(x,k_\perp^{\,2})$ and  $D^q(x_v,\gamma^0,v)$ in terms of 
manifestly Lorentz invariant amplitudes $A_i^q(P\cdot k,k^2)$, one would think it should 
be possible to carry out the last step of our proof, namely the demonstration of the 
equivalence of the representations in Eqs.~(\ref{Eq:def-quasi-III},~\ref{Eq:toward-Radyushkin-I}),
without the detour through the coordinate space representation of 
Ref.~\cite{Radyushkin:2017cyf}. We tried hard but have not been able to find such a 
direct proof which may indicate that exploring the coordinate space representation is 
an essential step.

\subsection{\boldmath Is the choice $\Gamma = \gamma^0$ or $\gamma^3$ preferable?}
\label{Subsec:which-choice-preferable}

Formulations in coordinate and momentum space give complementary insights. It 
may therefore be instructive to investigate the question which of the two choices is
preferable from the perspectives of the ${\cal M}^q(\nu,z^2)$ and ${\cal J}^q(\nu,z^2)$
amplitudes as well as the $A_i^q(P\cdot k,k^2)$ amplitudes. Let us begin with the coordinate
space representation. The expression for $D^q(x_v,\gamma^3,v)$ is obtained by choosing the
index $\mu=3$ in the correlator $C^\mu(P,z)$ in Eq.~(\ref{Eq:correlator-coordinate-space})
and $z^\mu=(0,\vec{0}_\perp,z^3)$ where $\nu = -P\cdot z = P^3z^3$ or $z^3=\frac{\nu}{P^3}$.
In analogy to Eq.~(\ref{Eq:M-QPDF-I}), the QPDF $D^q(x_v,\gamma^3,v)$ is defined by
\be
    C^3(P,z) = 2P^3 {\cal M}^q(\nu,-z_3^2)+2M^2z^3{\cal J}^q(\nu,-z_3^2) =
    2P^3 \int_{-\infty}^\infty dx_v\,e^{i\nu x_v} D^q(x_v,\gamma^3,v)
\ee
which can be inverted and simplified using  $z^3=\frac{\nu}{P^3}$ as follows
\ba\label{Eq:choices-in-coordinate-space-1}
     v\,D^q(x_v,\gamma^0,v)
     &=& \int_{-\infty}^\infty \frac{d\nu}{2\pi}\, e^{-i\nu x_v} \;{\cal M}^q(\nu,-z_3^2)
     \nonumber\\
     D^q(x_v,\gamma^3,v)
     &=& \int_{-\infty}^\infty \frac{d\nu}{2\pi}\, e^{-i\nu x_v} 
     \biggl({\cal M}^q(\nu,-z_3^2)+\frac{M^2}{P_3^2}\,\nu {\cal J}^q(\nu,-z_3^2)\biggr)
\ea
where we have included for comparison the result for $D^q(x_v,\gamma^0,v)$ discussed
already in Sec.~\ref{Sec:Radyushkin-formula}. Since $D^q(x_v,\gamma^3,v)$ contains
the power-suppressed term $\frac{M^2}{P_3^2}\,\nu {\cal J}^q(\nu,-z_3^2)$, 
the choice $\Gamma=\gamma^0$ appears preferable because in $D^q(x_v,\gamma^0,v)$ this
power-suppressed term is absent to begin with \cite{Radyushkin:2016hsy,Radyushkin:2017cyf}.

To investigate this question from a momentum space perspective, we need to determine the
explicit expressions for the amplitudes ${\cal M}^q(\nu,z^2)$ and ${\cal J}^q(\nu,z^2)$
from Eq.~(\ref{Eq:toward-Radyushkin-II}). The amplitude
$A_2^q(P\cdot k,k^2)$ accompanied by the four-vector $P^\mu$ is readily Fourier
transformed giving a contribution ${\cal M}^q_1(\nu,z^2)$ to ${\cal M}^q(\nu,z^2)$.
The amplitude $A_3^q(P\cdot k,k^2)$ is accompanied by the four-vector $k^\mu$ which is
the integration variable. Due to Lorentz invariance, the Fourier transform of this
term contributes to ${\cal M}^q(\nu,z^2)$ a second contribution ${\cal M}^q_2(\nu,z^2)$
and gives rise to the amplitude ${\cal J}^q(\nu,z^2)$ according to 
\be\label{Eq:correlator-coordinate-space-part-with-k-mu}
   2 \int d^4k\,e^{-ik\cdot z}\,k^\mu\,  A_3^q(P\cdot k,k^2) =
   P^\mu {\cal M}^q_2(\nu,z^2) + z^\mu M^2{\cal J}^q(\nu,z^2) \, , \qquad \nu = -P\cdot z\,.
\ee
Contracting the above equation with the four-vectors $P^\mu$ and $z^\mu$ yields two
equations for the two unknowns ${\cal M}^q_2(\nu,z^2)$ and ${\cal J}^q(\nu,z^2)$.
Solving these equations (and adding ${\cal M}^q_1(\nu,z^2)$ which is just the Fourier
transform of $A_2^q$) yields
\bsub\label{Eq:amplitudes-M-J}\ba
    {\cal M}^q(\nu,z^2) &=& 2\int d^4k\,e^{-ik\cdot z}\biggl[A_2^q-
    \frac{\;(k\cdot z)\,\nu+(P\cdot k)\,z^2\;}{\;\;\;\nu^2+M^2z^2}\,A_3^q\biggr]
    \\
    {\cal J}^q(\nu,z^2) &=& 2\int d^4k\,e^{-ik\cdot z}\biggl[\phantom{A_2^q} - 
    \frac{(P\cdot k)\,\nu+(k\cdot z)\,M^2}{M^2(\nu^2+M^2z^2)\;\;\;}\,A_3^q\biggr]
\ea\esub
where $A_i^q=A_i^q(P\cdot k,k^2)$. We make a first interesting observation.
The amplitude ${\cal J}^q(\nu,z^2)$ receives a contribution from $A_3^q$ just as
${\cal M}^q(\nu,z^2)$ does. Both $A_2^q$ and $A_3^q$ enter the description of the
twist-2 PDF $f_1^q(x)$ in QCD and in models. In the parton model (or Wandzura-Wilczek
approximation) where ``interaction dependent'' (so called tilde) terms are neglected,
the amplitude $A_2^q(P\cdot k,k^2)$ vanishes \cite{Aslan:2022wqc}. Thus, from this point
of view ${\cal J}^q(\nu,z^2)$ by itself, does not appear to be any lesser than
${\cal M}^q(\nu,z^2)$. It merely appears with a power-suppressed prefactor.

Let us now inspect the two choices $D^q(x_v,\Gamma,v)$ in
Eq.~(\ref{Eq:choices-in-coordinate-space-1}). Setting $z^\mu=(0,0,0,z^3)$ in
Eq.~(\ref{Eq:amplitudes-M-J}) with $z^3=\tfrac{\nu}{P^3}$ and choosing
$\Gamma=\gamma^0$ or $\gamma^3$ yields for $\nu$-integrands of the two
choices the following expressions in terms of $A_i$-amplitudes
\bsub\label{Eq:choices-in-coordinate-space-2}\ba
   \Gamma = \gamma^0: \;\qquad
   {\cal M}^q(\nu,-\tfrac{\nu^2}{P_3^2})\phantom{+\frac{M^2}{P_3^2}\,\nu {\cal J}^q(\nu,-z_3^2)} &=&
   2\int d^4k\,e^{i\nu\frac{k^3}{P^3}} \biggl[A_2^q
     + \frac{\frac{k^3}{P^3}+\red{\frac{P\cdot k}{P_3^2}}}{1+\red{\frac{M^2}{P_3^2}}}\,A_3^q\biggr]
   \label{Eq:choices-in-coordinate-space-2a}\\
   \Gamma = \gamma^3: \qquad
   {\cal M}^q(\nu,-\tfrac{\nu^2}{P_3^2})+\frac{M^2}{P_3^2}\,\nu {\cal J}^q(\nu,-z_3^2) &=&
   2\int d^4k\,e^{i\nu\frac{k^3}{P^3}} \biggl[
     \underbrace{A_2^q
       + \frac{\frac{k^3}{P^3}+\red{\frac{P\cdot k}{P_3^2}}}
              {1+\red{\frac{M^2}{P_3^2}}}\,A_3^q}_{{\cal M}^q(\nu,-\tfrac{\nu^2}{P_3^2})}
     +\underbrace{
       \frac{\frac{k^3}{P^3}\,\red{\frac{M^2}{P_3^2}}-\red{\frac{P\cdot k}{P_3^2}}}
            {1+\red{\frac{M^2}{P_3^2}}}\,A_3^q}_{\frac{M^2}{P_3^2}\nu{\cal J}^q(\nu,-\tfrac{\nu^2}{P_3^2})}
     \biggr]
     \label{Eq:choices-in-coordinate-space-2b}
\ea
The ratio $\frac{k^3}{P^3}$ corresponds to $x_v$ which asymptotically approaches $x$.
The frame-independent factor $P\cdot k$ appears always in a ratio with respect to
$P_3^2$ which makes it a power correction for $P^3\to\infty$. We also encounter power 
corrections of the type $\frac{M^2}{P_3^2}$. All these power-corrections are highlighted 
in color in Eq.~(\ref{Eq:choices-in-coordinate-space-2}).

Now we make the second remarkable observation. The case $\Gamma=\gamma^0$ in
Eq.~(\ref{Eq:choices-in-coordinate-space-2a}) stays as it is, while in the 
$\Gamma=\gamma^3$ case in Eq.~(\ref{Eq:choices-in-coordinate-space-2b}) all the
power corrections highlighted in color mutually compensate and we obtain
\be\label{Eq:choices-in-coordinate-space-2c}
   \Gamma = \gamma^3: \qquad
   {\cal M}^q(\nu,-\tfrac{\nu^2}{P_3^2})+\frac{M^2}{P_3^2}\,\nu {\cal J}^q(\nu,-z_3^2)
   =
   2\int d^4k\,e^{i\nu\frac{k^3}{P^3}} \biggl[A_2^q + \frac{k^3}{P^3}\,A_3^q\biggr]\,,
   \hspace{3cm}
\ee\esub
which is rather remarkable. Inserting the results from
Eq.~(\ref{Eq:choices-in-coordinate-space-2}) in
Eq.~(\ref{Eq:choices-in-coordinate-space-1}) and taking the 
$\nu$-integrals we obtain
\bsub\label{Eq:choices-in-coordinate-space-3}
\ba\label{Eq:choices-in-coordinate-space-3a}
     D^q(x_v,\gamma^0,v)
     &=&    2\int d^4k\,\delta\biggl(x_v-\frac{k^3}{P^3}\biggr)\,\biggl[A_2^q
       + \frac{x_v+\red{\frac{P\cdot k}{P_3^2}}}{1+\red{\frac{M^2}{P_3^2}}}\,A_3^q\biggr]\,
     \sqrt{1+\red{\tfrac{M^2}{P_3^2}}}
     \\
     \label{Eq:choices-in-coordinate-space-3b}
     D^q(x_v,\gamma^3,v)
     &=&   2\int d^4k\,\delta\biggl(x_v-\frac{k^3}{P^3}\biggr)\,\biggl[A_2^q
       + x_v\,A_3^q\biggr]
\ea\esub
The expressions in Eq.~(\ref{Eq:choices-in-coordinate-space-3}) coincide
with those in Eq.~(\ref{Eq:def-quasi-III}) as one can immediately see for
$D^q(x_v,\gamma^3,v)$ and, after some algebra, also for $D^q(x_v,\gamma^0,v)$.

Several remarks are in order. The square-root factor in
Eq.~(\ref{Eq:choices-in-coordinate-space-3a}) is $\frac1v$, cf.\ the end of 
Sec.~\ref{Sec:notation-Pz-vs-v}, and can be avoided by using the conventions of
Refs.~\cite{Radyushkin:2016hsy,Radyushkin:2017cyf}, see App.~\ref{App:conventions},
or by computing $v D^q(x_v,\gamma^0,v)$. The other power corrections
cannot be avoided. Since the practical goal is to study the QPDFs at very large $P^3 \gg M$ 
where the power corrections become small and QPDFs approach the PDF limit, there is a lot of 
practical interest to start with a representation of QPDFs which is more likely to 
exhibit faster convergence toward PDFs. Based on the coordinate space 
representation in Eqs.~(\ref{Eq:choices-in-coordinate-space-1}) 
the choice $\Gamma=\gamma^0$ appears more favorable because it avoids from the 
very beginning a power-correction term proportional to $J^q(\nu,-z_3^2)$.
However, revisiting the situation in momentum-space in terms of the
$A_i^q(P\cdot k,k^2)$ amplitudes reveals that the inclusion of the power-correction 
term proportional to ${\cal J}^q(\nu,-z_3^2)$ in the case of $\Gamma=\gamma^3$ has 
actually a positive side effect: it cancels out power corrections present in the term 
proportional to ${\cal M}^q(\nu,-z_3^2)$ which remain uncompensated in $D^q(x_v,\gamma^0,v)$, 
see Eqs.~(\ref{Eq:choices-in-coordinate-space-3}). This in turn suggests that 
$D^q(x_v,\gamma^3,v)$ might be a better candidate for  faster convergence 
toward PDFs. 

The power corrections which enter in 
$D^q(x_v,\gamma^0,v)$ in Eq.~(\ref{Eq:choices-in-coordinate-space-3a}) and are absent in
$D^q(x_v,\gamma^3,v)$ in Eqs.~(\ref{Eq:choices-in-coordinate-space-3b}) are of the type 
$\frac{P\cdot k}{P_3^2}$ or $\frac{M^2}{P_3^2}$ and can be referred to as a type of 
kinematical or target-mass corrections. In addition to this, there are of course 
other, dynamical power corrections due to higher twist contributions which are not
apparent in our approach. For instance, it is evident from the momentum sum rules, 
see Sec.~\ref{Sec:mom-sum-rule}, that $D^q(x_v,\gamma^3,v)$ contains a higher-twist  
contamination manifest through the EMT form factor $\bar{c}^q(0)$ which is twist-4
while the momentum sum rule of $D^q(x_v,\gamma^0,v)$ exhibits no such contamination. 
Nevertheless, as far as the simpler quark models are concerned, one may expect faster
convergence for $\Gamma=\gamma^3$ than $\gamma^0$, although this remains to be asserted
by explicit model calculations. 
For lattice fermion discretizations that break chiral symmetry, the $\gamma^3$ operator 
can mix with the scalar operator, whereas the $\gamma^0$ operator avoids this mixing and is 
therefore sometimes preferred in lattice-QCD calculations; see, e.g., 
Refs.~\cite{Constantinou:2017sej,Alexandrou:2017huk,Cichy:2018mum,Chen:2017mzz,Chen:2018xof,Constantinou:2020hdm}.

The results and considerations presented thus far are valid for any quark model which 
respects Lorentz symmetry. In the next section we shall choose a specific model to
illustrate and test the conclusions drawn in this section.

\newpage
\section{QPDF\lowercase{s} in the Covariant Parton Model}

To present practical results we choose in this work the 
Covariant Parton Model (CPM) based on a systematic extension
of Feynman's parton model concept \cite{Feynman:1969ej,Feynman:1973xc}
to the description of the partonic structure of hadrons~\cite{
Zavada:1996kp,Zavada:2001bq,Zavada:2002uz,Efremov:2004tz,
Zavada:2007ww,Efremov:2009ze,DAlesio:2009cps,Zavada:2009ska,
Efremov:2009vb,Efremov:2010mt,Zavada:2011cv,Zavada:2013ola,
Zavada:2015gaa,Bastami:2020rxn,Aslan:2022wqc,Aslan:2022kmd}.

\subsection{Quark correlator, amplitudes and PDFs in the CPM}
\label{Subsec:modelling-in-CPM}

Let us first briefly review the main results on the CPM 
\cite{Zavada:1996kp,Zavada:2001bq,Zavada:2002uz,Efremov:2004tz,
Zavada:2007ww,Efremov:2009ze,DAlesio:2009cps,Zavada:2009ska,
Efremov:2009vb,Efremov:2010mt,Zavada:2011cv,Zavada:2013ola,
Zavada:2015gaa,Bastami:2020rxn,Aslan:2022wqc,Aslan:2022kmd}
following the field-theoretic formulation of
\cite{Bastami:2020rxn,Aslan:2022wqc}. In the
CPM, the partons are on-shell and the amplitudes $A_i^q$ for $i=2,\,3$ 
needed for the unpolarized case are given by 
\ba
    A_2^q(P\cdot k,k^2) &=& 0 \, , \nonumber\\
    A_3^q(P\cdot k,k^2) &=& M\,\delta(k^2-m_q^2)\,
    \Theta_{\rm kin}(P\cdot k)\,
    {\cal G}^q(P\cdot k) \, ,\nonumber\\
    \Theta_{\rm kin}(P\cdot k) &=& 
    \Theta_{\rm kin}^{(+)}(P\cdot k)+\Theta_{\rm kin}^{(-)}(P\cdot k)
    \, , \quad
    \Theta_{\rm kin}^{(\pm)}(P\cdot k) = 
    \Theta(\pm P\cdot k)\,\Theta\left((P\mp k)^2\right)\,
    \label{Eq:CPM-Ai}
\ea
where the $\Theta$-functions ensure correct analytical properties
\cite{DAlesio:2009cps,Bastami:2020rxn} and the $\delta$-function contains 
the on-shell condition. 
The vanishing of the amplitude $A_2^q$ is a  prediction of the parton model, 
although the same was observed also in some models with interactions
\cite{Lorce:2014hxa}.
The amplitude $A_3^q$ is described in terms of the covariant function 
${\cal G}^q(P\cdot k)$ which for $P\cdot k>0$ describes unpolarized quarks. 
For $P\cdot k<0$ it describes antiquarks via the relation 
${\cal G}^{\bar q}(P\cdot k)={\cal G}^q(-P\cdot k)$
\cite{Aslan:2022wqc}. The covariant functions ${\cal G}^a(P\cdot k)$ 
determine the unpolarized PDFs $f_1^a(x)$ for $a=q,\,\bar q$.
In the nucleon rest frame and neglecting current quark mass effects, 
the unpolarized PDFs are given by
\be\label{Eq:CPM-f1}
    f_1^a(x) = 2\,\pi\,x\,M^2\int\limits_{\frac12|x|M}^{\frac12M}dk\;
    {\cal G}^a(Mk) \;. 
\ee
This one-to-one correspondence between ${\cal G}^a(P\cdot k)$ and  $f_1^a(x)$
can be inverted to determine the covariant functions from a chosen input PDF 
parametrization according to
\ba\label{Eq:determine-cov-functions}
      {\cal G}^a(P\cdot k)
      = - \frac{1}{\pi M^3}\,\,
      \biggl[\frac{d}{dx}\,\frac{f_1^a(x)}{x}\biggr] \Biggl|_{x = \frac{2 P\cdot k} {M^2}}\,,
      \quad a = q, \;\bar{q}.
\ea
For massless partons, the variable $P\cdot k$ is constrained as $0 < P\cdot k < \frac12 M^2$ 
due to the $\Theta$-functions in Eq.~(\ref{Eq:CPM-Ai}). The choice of the renormalization scale 
$\mu$ of the input PDF in Eq.~(\ref{Eq:determine-cov-functions}) is part of the modeling.
The scale $\mu$ must be chosen to be several GeV, high enough for the parton model
concept to be applicable. 
In an analogous way, it is possible to describe polarized PDFs and extend the 
approach to TMDs as well as subleading twist
\cite{Zavada:1996kp,Zavada:2001bq,Zavada:2002uz,Efremov:2004tz,
Zavada:2007ww,Efremov:2009ze,DAlesio:2009cps,Zavada:2009ska,
Efremov:2009vb,Efremov:2010mt,Zavada:2011cv,Zavada:2013ola,
Zavada:2015gaa,Bastami:2020rxn,Aslan:2022wqc,Aslan:2022kmd}.
Notice that $f_1^a(x)/x$ is monotonically decreasing for all flavors in all paramaterizations 
which implies that ${\cal G}^a(P\cdot k)$ is positive.

\subsection{Quasi PDF\lowercase{s} in the CPM}
\label{Sec:quasi-in-CPM}

It is convenient to derive first the CPM expression for the QPDF with 
$\Gamma=\gamma^3$ which is a simpler case than $\Gamma=\gamma^0$.
For that we insert the model results for the amplitudes (\ref{Eq:CPM-Ai}) 
into Eq.~(\ref{Eq:def-quasi-3-v-limit}), carry out the integration over 
$dk^0$ exploring $\delta(k^2-m_q^2)$, and neglect $m_q$ which is appropriate 
for current masses of light quarks. This yields the results
\ba
    D^q_{\pm}(x_v,\gamma^3,v)
     &=& 2\,x_v\,M \int \frac{d^3k}{2|\vec{k}|}\,
    \Theta(M-2|\vec{k}|)\,
    {\cal G}^q\bigl(\pm M|\vec{k}|\bigr)\,\delta\bigl(\pm x_v-\frac{k^3+v|\vec{k}|}{vM}\bigr) \,,
    \label{Eq:quasi-3}
\ea
where the $D^q_{\pm}(x_v,\gamma^3,v)$ denote the contributions  due to respectively 
$\Theta_{\rm kin}^{(\pm)}(P\cdot k)$ in Eq.~(\ref{Eq:CPM-Ai}) which need to be 
added up to the total QPDF, but for now it is convenient to distinguish them 
for pedagogical purposes. In Eq.~(\ref{Eq:quasi-3}) for $D^q_\pm(x_v,\gamma^3,v)$
we substituted $k^3\to (-k^3)$ under the $d^3k$ integral. 
Setting  $k^3=|\vec{k}|\,\cos\vartheta$ and $k=|\vec{k}|$ and integrating in
spherical coordinates over the angular variables $d\varphi \,d\cos\vartheta$,
we obtain
\be\label{Eq:D-CPM-3}
    D^q_{\pm}(x_v,\gamma^3,v) = 2\,\pi\,v\,x_v\,M^2 
    \int\limits_{L_\pm(v)}^{\frac12M}dk\;{\cal G}^q(\pm Mk) \,,
    \quad L_\pm(v) = \frac{v\,|x_v|\,M}{1\pm v\,{\rm sign}(x_v)}\,.
\ee
In the case $\Gamma=\gamma^0$ we obtain in the same way the results
\be\label{Eq:D-CPM-0}
    D^q_{\pm}(x_v,\gamma^0,v) = 2\,\pi\,M
    \int\limits_{L_\pm(v)}^{\frac12M}dk\;{\cal G}^q(\pm Mk)
    \biggl(x_vMv^2 \pm (1-v^2)\,k\biggr)\,.
\ee

The lower integration limits $L_\pm(v)$ in Eqs.~(\ref{Eq:D-CPM-3},~\ref{Eq:D-CPM-0}) 
arise from integrating the $\delta$-function in Eq.~(\ref{Eq:quasi-3}) over
$d \cos\vartheta$. The upper limit $\frac12M$ can be traced back to the step 
functions in Eq.~(\ref{Eq:CPM-Ai}). The latter also ensure that $L_\pm(v)<\frac12M$ 
and determine the support properties of QPDFs which we will discuss in the next 
section. The final model results for the QPDF are given by
\be\label{Eq:QPDF-CPM-total}
     D^q(x_v,\Gamma,v) = D^q_{+}(x_v,\Gamma,v) + D^q_{-}(x_v,\Gamma,v) \,.
\ee
In the limit $v\to 1$ when $x_v\to x$, the QPDFs approach 
\be\label{Eq:CPM-f1-limits}
      \lim\limits_{v\to1}D^q(x_v,\gamma^0,v)
    = \lim\limits_{v\to1}D^q(x_v,\gamma^3,v)
    = 2\,\pi\,x\,M^2\int\limits_{\frac12|x|M}^{\frac12M}dk\;
      {\cal G}^q({\rm sign}(x)Mk) 
    = f_1^q(x)\,,
\ee
Considering the relation ${\cal G}^q(-P\cdot k)={\cal G}^{\bar q}(P\cdot k)$ 
\cite{Aslan:2022wqc}, we see that at negative $x$ the Eq.~(\ref{Eq:CPM-f1-limits}) 
describes antiquark PDFs according to Eq.~(\ref{Eq:def-PDF-qbar}). 
The result in Eq.~(\ref{Eq:CPM-f1-limits}) agrees with the CPM expression 
for the unpolarized PDFs \cite{Efremov:2010mt}. Hence, the CPM results for 
QPDFs have the correct limit (\ref{Eq:general-limit-quasi-PDF}).

\subsection{Support properties of QPDF\lowercase{s} and illustration with toy model input}
\label{Sec:support-leaking}

It is interesting to investigate the support properties of QPDFs 
in the CPM in detail. As mentioned in Sec.~\ref{Sec:quasi-in-CPM}, 
the step functions in Eq.~(\ref{Eq:CPM-Ai}) ensure that the lower and upper 
integration limits in Eqs.~(\ref{Eq:D-CPM-3},~\ref{Eq:D-CPM-0}) 
satisfy $L_\pm(v)<\frac12M$. 
This condition determines the support properties of QPDFs.
The total QPDF, Eq.~(\ref{Eq:QPDF-CPM-total}), has non-zero support 
in a symmetric $x_v$ interval, while the individual contributions
$D^q_{\pm}(x_v,\Gamma,v)$ have support in asymmetric intervals as follows
\bsub\label{Eq:support}
\ba
    \label{Eq:support-QPDF}
    D^q(x_v,\Gamma,v) \neq 0 \quad \mbox{if} \quad
    -\,\frac{1+v}{2v} \le x_v \le  \frac{1+v}{2v}
    \quad \stackrel{v\to 1}{\xrightarrow{\hspace{1cm}}} \quad
    -1 < x < 1\,,
    \label{Eq:support-QPDF-plus} \\
    D^q_{+}(x_v,\Gamma,v) \neq 0 \quad \mbox{if} \quad
    -\,\frac{1-v}{2v} \le x_v \le  \frac{1+v}{2v} \, 
    \quad \stackrel{v\to 1}{\xrightarrow{\hspace{1cm}}} \quad
    \phantom{-}0 < x < 1\,, \\
    \label{Eq:support-QPDF-minus}
    D^q_{-}(x_v,\Gamma,v) \neq 0 \quad \mbox{if} \quad
    -\,\frac{1+v}{2v} \le x_v \le  \frac{1-v}{2v} 
    \quad \stackrel{v\to 1}{\xrightarrow{\hspace{1cm}}} \quad
    -1 < x < 0\,,
\ea\esub
where we indicate the limits $v\to 1$ when QPDFs become PDFs.
These support properties imply that the one-to-one correspondence 
between PDFs $f_1^a(x)$ and covariant distributions ${\cal G}^a(P\cdot k)$ 
for $a=q,\,\bar q$, cf.\ Eqs.~(\ref{Eq:CPM-f1},~\ref{Eq:determine-cov-functions}),
no longer holds for QPDFs. Instead, we find an interesting ``leaking'' phenomenon
in the following sense.

For $v<1$ the covariant distribution  
${\cal G}^q(P\cdot k)$ determines $D^q_+(x_v,\Gamma,v)$, 
while ${\cal G}^{\bar q}(P\cdot k)$ determines $D^q_-(x_v,\Gamma,v)$. 
But the $D^q_\pm(x_v,\Gamma,v)$ contribute to both quark and antiquark 
QPDFs. As a consequence, the effects of the covariant quark distribution 
${\cal G}^q(P\cdot k)$ (which for $v\to1$ is solely responsible for 
the quark PDF) spill over into the antiquark region for $v<1$.
And vice versa, the effects of ${\cal G}^{\bar q}(P\cdot k)$
(which for $v\to1$ is solely responsible for the antiquark PDF) 
leak into the antiquark region.

This leaking reflects the fact that a  probabilistic parton density
interpretation, and hence distinction of quarks and antiquark degrees
of freedom, requires the infinite momentum frame. In frames with $v<1$ 
the distinction between quarks and antiquarks is lost.
(In lightcone quantization \cite{Brodsky:1997de} a probabilistic 
interpretation can be given for any $v$.)
When a nucleon moves with a velocity less than the speed of light 
(needed for a rigorous partonic interpretation in the infinite momentum frame),
there are quark configurations inside the nucleon state which move in the opposite 
direction and ``mimic'' antiquarks \cite{Diakonov:1997vc}. 
This corresponds in the CPM to the ``leaking'' of $D^q_+(x_v,\Gamma,v)$ into negative $x_v$. 
Similarly, antiquark configurations exist for $v<1$ in the nucleon state which move in 
the opposite direction to nucleon, mimicking quarks \cite{Diakonov:1997vc}. 
This corresponds in the CPM to the ``leaking'' of $D^q_-(x_v,\Gamma,v)$ into positive $x_v$. 

%========== BEGIN FIGURE 1 =======================================
%
\begin{figure}[t!]
  \includegraphics[width=.3\linewidth]{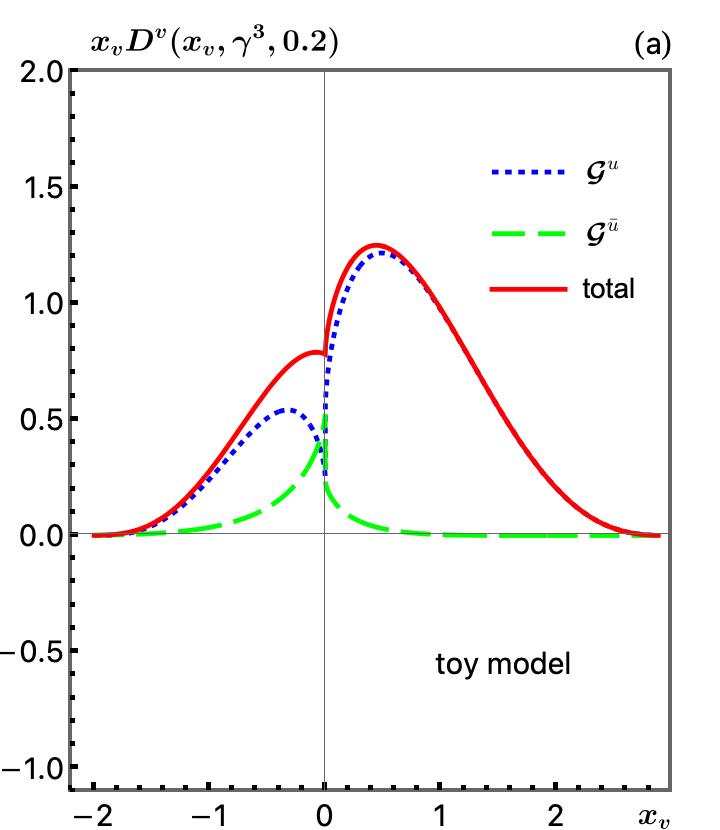}
  \hspace{1cm}
  \includegraphics[width=.3\linewidth]{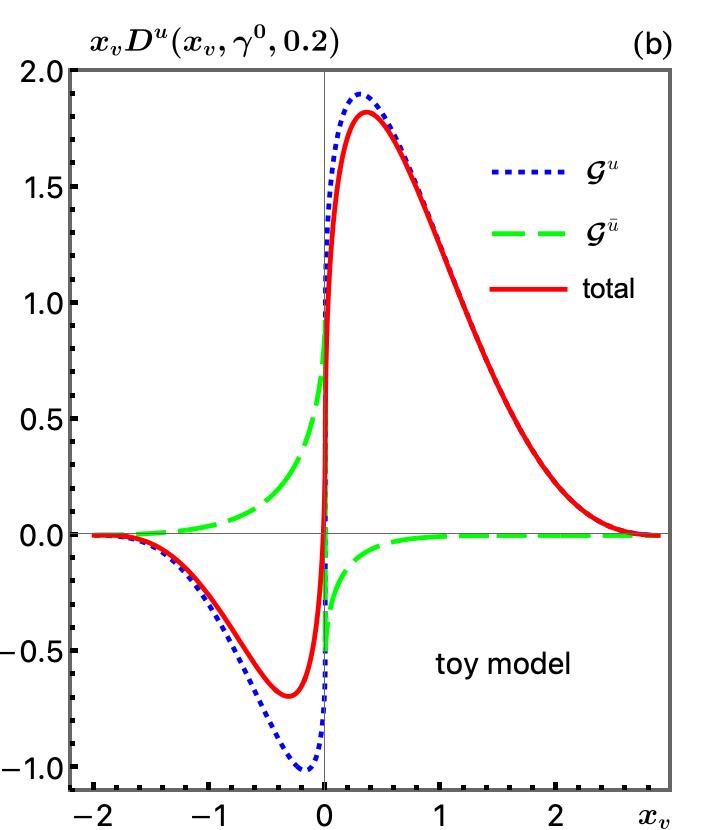}
\caption{\label{Fig1:leaking-in-toy-model} The QPDFs for 
(a) the case $\Gamma=\gamma^3$ and 
(b) the case $\Gamma=\gamma^0$ as functions of $x_v$ for $v=0.2$ 
in the CPM based on toy model input for pedagogical purposes as described
in text. For $v=1$ the covariant functions ${\cal G}^q(P\cdot k)$ and 
${\cal G}^{\bar q}(P\cdot k)$ describe respectively the quark PDF and 
antiquark PDFs. For $v<1$ the effects of quarks described in terms of 
${\cal G}^q(P\cdot k)$ ``leak'' into the antiquark region, and the
effects of antiquarks described by ${\cal G}^{\bar q}(P\cdot k)$ 
``leak'' into the quark region.}
\end{figure}
%
%============ END FIGURE 1 =======================================

In Fig.~\ref{Fig1:leaking-in-toy-model} we illustrate the ``leaking phenomenon'' 
using toy model PDFs as input 
which behave like $f_1^a(x)\propto 1/x$ at small $x$. This has the advantage that
$D^a(x_v,\Gamma,v)\propto 1/x_v$ at small $x_v$ such that there is no singularity 
at $x_v=0$ when plotting $x_vD^a(x_v,\Gamma,v)$ over the entire $x_v$ interval.
The toy model PDFs are used only in this section to demonstrate generic features. 
The small-$x_v$ behavior of QPDFs in the CPM for 
any input PDFs will be discussed in Sec.~\ref{Sec:small-xv} and numerical results 
from realistic parametrizations will be shown in Sec.~\ref{Sec:numeric}.

For pedagogical purposes, we focus on $u$-flavor and determine the CPM covariant 
functions ${\cal G}^a(P\cdot k)$ in Eq.~(\ref{Eq:determine-cov-functions}) from 
analytical toy model PDFs defined as
$f_1^a(x) = A_a\,x^{-B_a}(1+C_a\,x^{1/2}+D_a\,x)\,(1-x)^{E_a}$ 
with the parameters
$A=A_u=A_{\bar u}=0.3$,
$B=B_u=B_{\bar u}=1.0$,
$C_u=3.0$,  $C_{\bar u}=-1.6$,
$D_u=10.0$, $D_{\bar u}=1.25$,
$E_u=3.0$,  $E_{\bar u}=5.0$. 
This toy model is not unrealistic and chosen such that for 
$0.01\lesssim x\lesssim 0.7$ it approximates the NLO LHAPDF 
parametrization from \cite{Hou:2019qau} at the scale $\mu^2=4\,{\rm GeV}^2$ 
within few percent accuracy and accurately reproduces the first two 
Mellin moments. Notice that the parameters $A=A_a$ and $B=B_a<2$ must be the 
same for $a=q,\;\bar{q}$ for the sum rules to be convergent
(as is the case also for realistic PDFs). 
For $x\lesssim 10^{-2}$ we deliberately choose to deviate from realistic PDFs 
for pedagogical reasons, and the large-$x$ region is of no relevance for the 
purposes of this section. 

For the relatively low nucleon velocity $v=0.2$ chosen in 
Fig.~\ref{Fig1:leaking-in-toy-model} which corresponds to $P_z=0.19\,{\rm GeV}$,
the support of $D^q_+(x_v,\Gamma,v)$ is the region $-2\le x_v\le 3$,
while that of $D^q_-(x_v,\Gamma,v)$ is the region $-3\le x_v\le 2$.
It is a general feature of the CPM that for lower $v$, the QPDFs 
spread further beyond the PDF support region $|x|<1$, 
see Eqs.~(\ref{Eq:support-QPDF}-\ref{Eq:support-QPDF-minus}).
(Another general feature worth mentioning in passing is that QPDFs 
diverge in the opposite limit $v\to0$.)

In Fig.~\ref{Fig1:leaking-in-toy-model}a we see that for low velocities $D^q_+(x_v,\gamma^3,v)$ 
(which would provide the sole contribution to $f_1^q(x)$ for $v\to1$) 
leaks so significantly into the negative $x_v$ such that actually $D^q_+(x_v,\gamma^3,v)$ 
is the dominant contribution in this region. The effect is mutual 
and $D^q_-(x_v,\gamma^3,v)$ (responsible for the sole contribution to the antiquark PDF 
for $v\to1$) leaks into positive $x_v$ but has there a much smaller effect and constitutes
a modest correction to the total QPDF. The contributions of 
$x_vD^q_\pm(x_v,\gamma^3,v)$ to the total QPDF in all $x_v$-regions are always positive 
and add up. 

In Fig.~\ref{Fig1:leaking-in-toy-model}b in the case $\Gamma=\gamma^0$ we also observe 
a significant leaking of $x_v D^q_+(x_v,\gamma^0,v)$ into the negative $x_v$-region, and 
less pronounced leaking of $x_v D^q_-(x_v,\gamma^0,v)$ into positive $x_v$. 
But the leaking pattern is fundamentally 
different in the two cases: for $\Gamma=\gamma^0$ the respective ``leaking-in'' contributions 
have opposite signs and lead to cancellations, see Fig.~\ref{Fig1:leaking-in-toy-model}b. 
In our case (which is realistic for that matter) at lower velocities the cancellation in the 
negative $x_v$-region is so strong that it reverses the sign of the QPDF.
Whether or not in the case of $\Gamma=\gamma^0$ the leaking of $x_vD^q_+(x_v,\gamma^0,v)$
into the negative $x_v$-region is strong enough for $x_vD^q_+(x_v,\gamma^0,v)$ to
dominate over $x_vD^q_-(x_v,\gamma^0,v)$ and reverse the sign of the total QPDF, 
this depends on the input PDFs. This is generally the case for PDFs where $xf_1^q(x)$ 
has a pronounced maximum in the valence $x$-region as is the case with more or less 
realistic $u$- and $d$-flavor PDFs as input.\footnote{\label{footnote-sea-type-QPDF} In 
    a less realistic
    toy model $f_{\rm toy}^a(x)=A\,x^{-1}(1-x)^{B_a}$ with $B_q=4$ and $B_{\bar{q}}=8$
    with $A$ fixed by the flavor number sum rule, the resulting leaking of 
    $x_vD^q_+(x,\gamma^0,v)$ does not overwhelm $x_vD^{q}_-(x,\gamma^0,v)$ for 
    $x_v<0$, and the total $x_vD^{q}(x,\gamma^0,v)$ is throughout positive.
    The reason why in this case the leaking pattern is unlike in 
    Fig.~\ref{Fig1:leaking-in-toy-model}b is because in this toy model 
    $xf_{\rm toy}^q(x)$ has no maximum. 
    The reversal of the sign of $x_vD^{q}(x,\gamma^0,v)$ at $x_v<0$ for low and
    moderate velocities requires a pronounced valence behavior of $xf_1^q(x)$.}
As $v$ increases, the magnitude of the sign-reversing 
leaking-in contribution to the antiquark QPDF decreases.

\subsection{Sum rules for QPDF\lowercase{s} in the CPM}

In order to prove that the sum rules in the CPM, we first derive the model
expressions for the form factors $F_1^q(0)$, $A^q(0)$,
$\bar{c}^q(0)$. Evaluating the general quark model 
expressions for these form factors derived in the 
Appendices \ref{App:em-FF} and \ref{App:EMT-FF} we 
obtain the results
\begin{align}\label{Eq:CPM-F1q}
    F_1^q(0)
    = &
    2 \int d^4k\;\frac{E_q}{M}\,A_3^q \,, \quad
    E_q=\sqrt{\vec{k}{ }^2+m_q^2} \\
    \label{Eq:CPM-EMT-Aq}
    A^q(0) 
    = &
    2 \int d^4k\;\biggl(\phantom{-}
    \frac{4}{3}\,\frac{\vec{k}{ }^2}{M^2}
    +\frac{m_q^2}{M^2}\biggr)\,A_3^q
    \,,
    \\
    \label{Eq:CPM-EMT-cbarq}
    \bar{c}^q(0)
    = & 
    2 \int d^4k\,\biggl(
    - \frac{1}{3}\;\frac{\vec{k}{ }^2}{M^2} \biggr)\,A_3^q\,,
\end{align}
where $E_q$ is the energy of the onshell partons and we keep track of current quark 
mass effects. 

Interestingly, in the CPM the form factors $A^q(0)$ and $\bar{c}^q(0)$ are related 
modulo current quark mass effects. By exploring the $\delta$-function $\delta(k^2-m_q^2)$ 
inherent in $A_3^q$ in Eq.~(\ref{Eq:CPM-Ai}), we obtain from 
Eqs.~(\ref{Eq:CPM-EMT-Aq},~\ref{Eq:CPM-EMT-cbarq}) 
\be
    A^q(0) = -4\,\bar{c}^q(0) +
    \frac{2m_q^2}{M^2}\int d^4k\;A_3^q \,.
\ee
For light quarks $m_q^2\ll M^2$, the CPM practically predicts the relation 
\be\label{Eq:predict-cbar-CPM}
    \bar{c}^q(0)=-\tfrac14\,A^q(0) \quad \mbox{(CPM).}
\ee
This is a benchmark result in the sense that Eq.~(\ref{Eq:predict-cbar-CPM}) 
is based purely on Lorentz symmetry and free field~theory.  
The CPM has been argued \cite{Boussarie:2023izj} to be practically equivalent 
to the Wandzura-Wilczek (WW) approximation in QCD where quark-gluon 
correlators are systematically neglected with respect to quark correlators,
see e.g.\ \cite{Bastami:2018xqd}. Hence our result Eq.~(\ref{Eq:predict-cbar-CPM})
can be more broadly viewed as a prediction of the WW approximation 

When interactions are present, one can in general expect deviations from
(\ref{Eq:predict-cbar-CPM}), and it is instructive to compare to results from other 
approaches. Remarkably, the relation (\ref{Eq:predict-cbar-CPM}) holds 
exactly in the bag model like in the CPM (provided $m_q=0$ is set in both cases)
albeit the bag model result refers to a low hadronic scale
\cite{Ji:1997gm,Neubelt:2019sou,Avakian:2010br}. Interestingly, in the bag model there is a 
``gluon'' contribution (mimicked by the bag) given by 
$\bar{c}^g(t) = - \sum_q \bar{c}^q(t)$ with
$\bar{c}^g(0) = \frac14$ for $m_q=0$ \cite{Neubelt:2019sou}.
The reason why (\ref{Eq:predict-cbar-CPM}) holds in the bag model may
presumably be due to the quarks inside the bag obeying the free Dirac 
equation, as they do in the CPM, but this point may deserve more model studies.
The chiral quark soliton model offers a different scenario where the quarks obey 
a Dirac equation coupled to a strong mean (``soliton'') field. This leads to a very 
different prediction, namely $\bar{c}^{u+d}(0)=0$ \cite{Goeke:2007fp} in distinction 
to (\ref{Eq:predict-cbar-CPM}). Notice that the separate $u$- and $d$-flavor 
results are non-zero with $\bar{c}^u(0)=-\bar{c}{ }^{d}(0)=0.04$
at a low scale $\mu^2\sim 0.4\,{\rm GeV}^2$ \cite{Won:2023rec}.
The instanton vacuum model predicts $\bar{c}^{u+d}(0)=0.014$ also at  
$\mu^2\sim 0.4\,{\rm GeV}^2$ \cite{Polyakov:2018exb}. We note these results refer
to low scales while the CPM is defined at higher scales so the above comparison
is qualitative.

For a more quantitative comparison, we resort to results from literature 
referring to higher scales \cite{Lorce:2017xzd,Liu:2021gco} where respectively the value 
$\sum_q \bar{c}^q(0) = - 0.11$ was obtained phenomenologically \cite{Lorce:2017xzd}  
and the value $-0.12$ from the lattice QCD calculation \cite{Liu:2021gco}. These results
refer to summation over the light flavors $q=u,\,d,\,s$ and the renormalization scale 
$\mu^2 = 4\,{\rm GeV}{ }^2$. % in $\overline{MS}$ scheme.
Choosing as input for the CPM the NLO LHAPDFA parametrization \cite{Hou:2019qau} 
at $\mu^2=4\,{\rm GeV}^2$ where
$A^u(0) = 0.347$,    % $A^u(0) = 0.347381$,
$A^d(0) = 0.195$,    % $A^d(0) = 0.195466$,
$A^s(0) = 0.032$,    % $A^s(0) = 0.0315323$,
we obtain the estimate 
$\sum_q\bar{c}^q(0) = -0.14$ which agrees within 20$\,\%$
with \cite{Lorce:2017xzd,Liu:2021gco}, i.e.\ 
for this property and at this scale the WW-approximation works reasonably 
well. This is indeed within the accuracy of the WW approximation in QCD as inferred 
phenomenologically for the twist-3 PDF $g_T^a(x)$ \cite{Accardi:2009au}. 

At first glance, one might hope to observe a better agreement at higher scales 
where the parton model picture is better justified (we recall that the choice of 
scale is part of modeling in the CPM, see Sec.~\ref{Subsec:modelling-in-CPM}).
Interestingly, this is not the case. In Ref.~\cite{Hatta:2018sqd} 
the QCD evolution properties of $\bar{c}^q(0)$ were studied and the 
asymptotic limit was derived 
(the notation is $\bar{c}^q(t,\mu^2)$ but throughout we omit 
the scale dependence for brevity and the result below refers to $t=0$)
\be\label{Eq:QCD-asymptotic-cbar}
    \lim\limits_{\mu^2\to\infty}\sum_q \bar{c}^q(0) = -\,\frac14\,
    \lim\limits_{\mu^2\to\infty}\sum_q A^q(0) - \frac{N_f}{6\,\beta_0}
\ee
where $\lim_{\mu^2\to\infty}\sum_q A^q(0) = N_f/(N_f+4C_F)$ with $C_F = (N_c^2-1)/(2N_c)$.
Here $N_c$ is the number of colors in QCD and $\beta_0=\frac{11}{3}C_A-\frac23\,N_f$
is the leading coefficient of the QCD $\beta$-function and $C_A=N_c$.
For $N_f = 3$ flavors the asymptotic QCD formula (\ref{Eq:QCD-asymptotic-cbar}) yields 
$\lim_{\mu^2\to\infty}\sum_q \bar{c}^q(0)=-0.15$ while the CPM yields $-0.09$. 
 The discrepancy is due to the piece $-N_f/(6\,\beta_0)$ in
Eq.~(\ref{Eq:QCD-asymptotic-cbar}) which arises from the trace anomaly in QCD.
It is not surprising to miss out on effects of quantum anomalies in a parton model 
framework where one neglects (quantum) interactions in first place. 
Still, the asymptotic CPM and QCD results agree within $40\,\%$. 
Combining with what we learned in the previous paragraph from the comparison to
Refs.~\cite{Lorce:2017xzd,Liu:2021gco}, we conclude that the CPM prediction for 
$\bar{c}^q(0)$ works within $(20-40)\,\%$. Incidentally, this is an ``accuracy''
one often observes in model studies.
The numerical values of $A^q(0)$ and $\bar{c}^q(0)$ are of 
importance for the decomposition of hadron masses in QCD 
\cite{Ji:1994av,Metz:2020vxd,Ji:2021qgo,Lorce:2021xku}.

\subsection{Further general properties of QPDF\lowercase{s} in the CPM}
\label{Sec:relation-D3-D0}

In this section, we discuss several general properties of QPDFs
in the CPM. One interesting general feature in the CPM is that 
$D^q(x_v,\gamma^3,v)$ and $D^{\bar q}(x_v,\gamma^3,v)$ are always 
positive for all velocities $0 < v \le 1$, i.e.\ the CPM predicts 
\be\label{Eq:positivity-D3}
      D^a(x_v,\gamma^3,v) \ge 0\,, 
      \quad a = q,\;\bar{q}, 
      \quad 0 < v \le 1\,.
\ee
This is true in the CPM for any input PDF, i.e.\ it is not specific 
to using toy model or realistic PDFs as input. The result in 
Eq.~(\ref{Eq:positivity-D3}) follows from the positivity of the covariant 
functions ${\cal G}^a(P\cdot k)$ for $a=q,\,\bar{q}$ which follows from the 
positivity of $f_1^a(x)$ via Eq.~(\ref{Eq:determine-cov-functions}).
Notice that the positivity of $f_1^a(x)$ is obvious in the parton model 
framework or at tree level in QCD, but cannot be proven in general in QCD 
\cite{Collins:2021vke}. Within the simpler framework of the CPM, the positivity 
of $D^a(x_v,\gamma^3,v)$ in Eq.~(\ref{Eq:positivity-D3}) is a rigorous result.

Another interesting generic property in the CPM is that the contributions 
$D^q_\pm(x_v,\Gamma,v)$ to the two representations for $\Gamma = \gamma^0$ and 
$\Gamma=\gamma^3$ are related to each other according to
Eqs.~(\ref{Eq:D-CPM-3},~\ref{Eq:D-CPM-0}) as
\ba
    D^q_\pm(x_v,\gamma^0,v) 
    &=& v\,D^q_\pm(x_v,\gamma^3,v)
    + \Delta D^q_\pm(x_v,v) \nonumber\\
    \Delta D^q_\pm(x_v,v) 
    &=& \pm (1-v^2)\;2\,\pi\,M\!\!
    \int\limits_{L_\pm(v)}^{\frac12M}dk\;k\,
    {\cal G}^q\left(\pm Mk\right)\,,
    \label{Eq:D0-D3-relation}
\ea
and the corresponding total QPDFs are related analogously according to 
Eq.~(\ref{Eq:QPDF-CPM-total}). This relation is very interesting. It shows 
that $D^q(x_v,\gamma^0,v)$ contains the expression for $D^q(x_v,\gamma^3,v)$ 
``diluted'' by an additional power of nucleon velocity and ``contaminated'' 
by the extra term $\Delta D_\pm^q(x_v,v)$ proportional to $(1-v^2)$. 
Based on the model expressions alone, it is not possible to tell which choice,
$\Gamma=\gamma^0$ or $\gamma^3$, may converge faster and we have to postpone 
this question until the numerical study.

The ``extra term'' $\Delta D_\pm^q(x_v,v)$ in Eq.~(\ref{Eq:D0-D3-relation})
is interesting for several reasons. First, it is responsible that there is 
no positivity property for $D^q_\pm(x_v,\gamma^0,v)$ at $v<1$ unlike for
$D^q_\pm(x_v,\gamma^3,v)$ in Eq.~(\ref{Eq:positivity-D3}). Second, it 
causes a different leaking behavior in the cases $\Gamma=\gamma^0$ and 
$\Gamma=\gamma^3$ as we observed in Sec.~\ref{Sec:support-leaking}.
If we read Eq.~(\ref{Eq:D0-D3-relation}) the other way round as
$v\,D^q_\pm(x_v,\gamma^3,v) = D^q_\pm(x_v,\gamma^0,v)-\Delta D^q_\pm(x_v,v)$,
then we recognize a third feature, namely $\Delta D_\pm^q(x_v,v)$ is 
responsible for the contribution of $\bar{c}^q(t)$ in the sum rule in 
Eq.~(\ref{Eq:general-mom2-quasi-3}) which is genuinely twist-4. 
In fact, this term comes with the prefactor 
$(1-v^2) = M^2/(M^2+P_3^2)$ which corresponds for $P^3\gg M$ to a
power suppression of ${\cal O}(\frac{M^2}{P_3^2})$. 

Notice that ``reading-Eq.~(\ref{Eq:D0-D3-relation})-in-another-way'' above was
necessary to reconcile with the general sum rule (\ref{Eq:general-mom2-quasi-3}).
But in the model this ``twist-4'' piece actually explicitly appears as additional 
term in $D^q_\pm(x_v,\gamma^0,v)$ and not in $D^q_\pm(x_v,\gamma^3,v)$ which is another
interesting observation. Where the power corrections appear may depend on the point
of view, see the discussion of power corrections in coordinate vs momentum space in 
Sec.~\ref{Subsec:which-choice-preferable}.
It would be very interesting to investigate this point further in other models.

In lattice QCD calculations, the mixing with twist-4 contributions
makes the studies of $D^q(x_v,\gamma^3,v)$ more involved compared to
$D^q(x_v,\gamma^0,v)$. In the CPM, all twists are described on equal footing
(within the WW-type approximation) and there is a priori no preference 
to favor $\Gamma=\gamma^0$ or $\Gamma=\gamma^3$.

\subsection{\boldmath Small-$x_v$ behavior of QPDF\lowercase{s} in the CPM}
\label{Sec:small-xv}

The small-$x$ behavior of PDFs determines the small-$x_v$ behavior of QPDFs.  
The connection can be derived analytically in the CPM as follows. Let a PDF have 
the small-$x$ behavior
\be\label{Eq:small-x-PDF}
    f_1^a(x) = A\,x^{-B} 
    + \dots \quad \mbox{for} \quad x \to 0\,,
    \quad a = q,\;\bar{q},
\ee
where the dots denote subleading terms. The constants $A$ and $B$ are positive 
and scale- and flavor-dependent which we do not indicate for notational brevity.
Based on Eq.~(\ref{Eq:determine-cov-functions}) the covariant function exhibits 
the behavior 
\be
    {\cal G}^a(P\cdot k) = \frac{A(B+1)}{\pi\,M^3}\,
    \biggl(\frac{2\,P\cdot k}{M^2}\biggr)^{\!\!-B-2} 
    + \dots \quad \mbox{for} \quad P\cdot k \to 0 \, ,
\ee
i.e the covariant functions diverge strongly for $P\cdot k\to 0$. 
These properties imply for the QPDFs the small-$x_v$ behavior 
\ba\label{Eq:small-x-quasi-1}
    D^a(x_v,\Gamma,v) =  C(\Gamma,v)\, A \, x_v^{-B} 
    + \dots  \quad \mbox{for} \quad x_v \ll \frac{1-v}{2v} \,,
\ea
with the velocity dependent coefficients 
(for $0<v\le 1$ and $B\neq 0$, for realistic parametrizations $1<B<2$)
\ba\label{Eq:small-x-quasi-2}
    C(\gamma^3,v) &=& 
    v\,\Biggl[\biggl(\frac{1+v}{2v}\biggr)^{B+1}+\biggl(\frac{1-v}{2v}\biggr)^{B+1}\Biggr]\,,
    \nonumber\\
    C(\gamma^0,v) &=& 
    v\,C(\gamma^3,v) + (1-v^2)\,\frac{B+1}{2B}
    \Biggl[\biggl(\frac{1+v}{2v}\biggr)^B-\biggl(\frac{1-v}{2v}\biggr)^B\Biggr] \,.
\ea
Both coefficients have the property $\lim_{v\to1}C(\Gamma,v)=1$ and we 
recover the correct small-$x$ behavior (\ref{Eq:small-x-PDF}) of the PDF
for $v\to1$. The results 
(\ref{Eq:small-x-quasi-1},~\ref{Eq:small-x-quasi-2}) can 
be viewed as predictions of the WW-type approximation.

\subsection{\boldmath Large-$x_v$ behavior of QPDF\lowercase{s} in the CPM}
\label{Sec:large-xv}

Also the large-$x_v$ behavior of QPDFs can be derived analytically in the CPM.
Let us recall, cf.\ Eq.~(\ref{Eq:support}), that in the CPM the QPDFs vanish 
for $x_v \ge x_{v, \rm max}$ with the definition
\be
     x_{v, \rm max} = \frac{1+v}{2v}
\ee
which is larger than unity for $0<v<1$ and  $\lim_{v\to 1} x_{v, \rm max} = 1$. 
This is notably different from QCD and the models we are aware of where the 
support of QPDFs extends to infinity. 
For $x_v\to x_{v,\rm max}$ the behaviour of the QPDFs in the CPM 
can be derived analytically analogously to Sec.\ref{Sec:small-xv}.
Let the large-$x$ behaviour of a PDF be given by
\be\label{Eq:large-x-PDF}
    f_1^a(x) = c_L\,(1-x)^N 
    + \dots \quad \mbox{for} \quad x \to 1\,, 
\ee
where the dots denote subleading terms. 
The constants $c_L$ and $N$ differ for each parton species 
$a = u, \,\bar{u}, \,d, \, \bar d, \dots$ and are scale-dependent and positive.
The covariant functions ${\cal G}^a(P\cdot k)$ vanish (for massless partons) 
for $2\,P\cdot k \ge M^2$ based on the constraints encoded in the $\Theta$-functions 
in Eq.~(\ref{Eq:CPM-Ai}), and the large $x$-behavior of the PDFs implies that
\be
    {\cal G}^a(P\cdot k) = \frac{c_L\,N}{\pi\,M^3}\,
    \biggl(1-\frac{2P\cdot k}{M^2}\biggr)^{N-1} + \; \dots 
    \quad \mbox{for} \quad 2\,P\cdot k \to M^2\,.
\ee
This in turn implies for QPDFs the large-$x_v$ behavior
(note that $v x_v$ below approaches $\frac12(1+v)+\dots$
for $x_v\to x_{v,\rm max}$)
\be\label{Eq:zzzzz}
    D^a(x_v,\Gamma,v)=
    c_L\,v\,x_v \,\left(1-\frac{x_v}{x_{v,\rm max}}\right)^N+\dots \; ,
    \quad \Gamma = \gamma^0,\;\gamma^3\,.
\ee
Interestingly, both choices $\Gamma = \gamma^0,\;\gamma^3$ exhibit the same large-$x_v$ 
behavior and are larger than the PDF at large $x_v$. The same behavior was observed in 
prior model studies \cite{Son:2019ghf, Bhattacharya:2018zxi, Bhattacharya:2019cme, Braun:2018brg}.

\newpage
\section{Numerical results for QPDF\lowercase{s} in CPM}
\label{Sec:numeric}

\noindent
\begin{wraptable}[10]{r}{98mm}
\centering
\begin{tabular}{l|llllll}
nucleon velocity $v$                    & \ \ 0.2  & \ \ 0.5  & \ \ 0.7  & \ \ 0.9  & \ \ 1         \\ \hline
nucleon momentum $P_z$ [GeV]            & \ \ 0.19 & \ \ 0.54 & \ \ 0.92 & \ \ 1.94 & \ \ $\infty$  \\ \hline
range of QPDF support $x_{v,\rm max}$ \ & \ \ 3    & \ \ 1.5  & \ \ 1.21 & \ \ 1.06 & \ \ 1
\end{tabular}
\caption{Nucleon velocities $v$ and momenta $P_z$ chosen for displaying 
numerical results along with the QPDF support $0\le x_v \le x_{v,\rm max}$ 
in the CPM ($v=0.2$ is used only in Fig.~\ref{Fig1:leaking-in-toy-model} 
and for pedagogical purposes). \label{Tab1}}
\end{wraptable}

We determine the covariant functions 
${\cal G}^a(P\cdot k)$ in Eq.~(\ref{Eq:determine-cov-functions}) using the 
NLO LHAPDF parametrization \cite{Hou:2019qau} at $\mu^2=4\,{\rm GeV}^2$ for
$a=u,\,d,\,\bar{u},\,\bar{d},\,s$ and compute the CPM results for 
$D^a(x_v,\Gamma,v)$ based on Eqs.~(\ref{Eq:D-CPM-3}--\ref{Eq:QPDF-CPM-total}).
Since $f_1^s(x)=f_1^{\bar s}(x)$ in the parametrization \cite{Hou:2019qau}, 
also the pertinent QPDFs coincide and we show results for only one of them.

In order to demonstrate the convergence of the QPDFs we choose the velocities 
$v=0.5,\,0.7,\,0.9$ which correspond to nucleon momenta $P_z$ as shown in 
Tab.~\ref{Tab1}. For these velocities, in the CPM the QPDFs have non-zero
support for $0 < x_v < x_{v,\rm max}$ with $x_{v,\rm max}$  as shown in 
Tab.~\ref{Tab1}.
Unlike in Fig.~\ref{Fig1:leaking-in-toy-model}, the antiquark PDFs will be
displayed for positive $x_v$ (recall that Eq.~(\ref{Eq:def-PDF-qbar})
allows one to switch back and forth between these equivalent displays).

\subsection{CPM results for \boldmath $x_vD^a(x_v,\Gamma,v)$}

\begin{figure}[b!]
\begin{tabular}{lll}  
      \includegraphics[width=.27\linewidth]{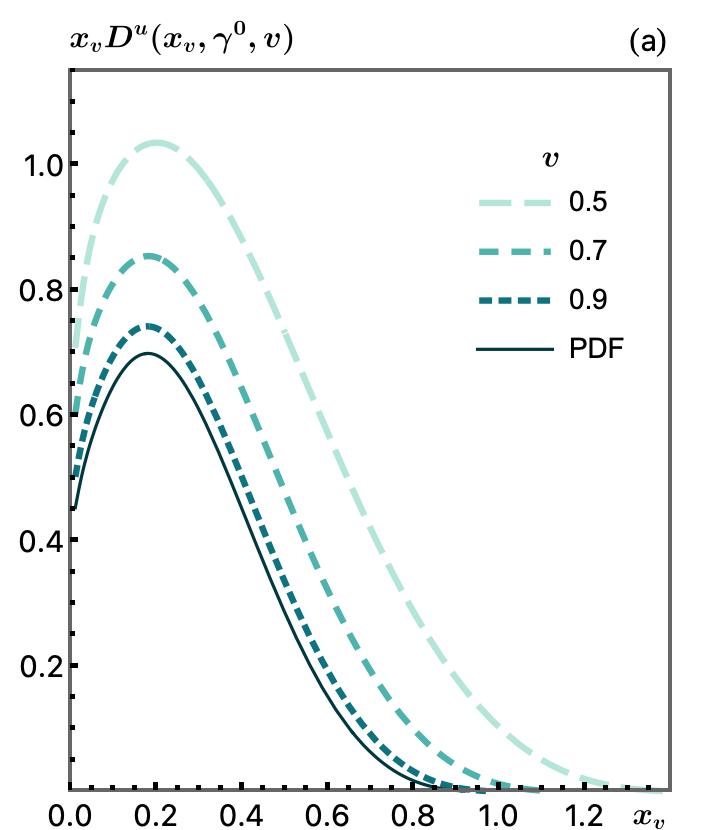}
    & \includegraphics[width=.27\linewidth]{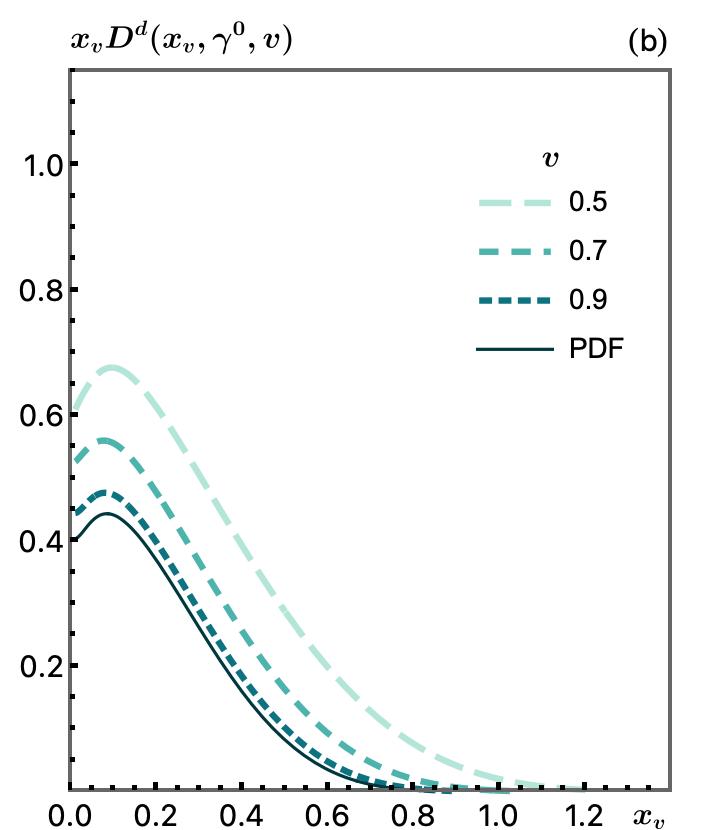}
    &   \hspace{3mm} 
        \begin{minipage}[b]{.32\linewidth}\vspace*{0pt} 
            \caption{\label{Fig-02:QPDF-gamma-0} The QPDFs $x_v D^a(x_v,\gamma^0,v)$ 
            in the CPM as functions of $x_v\ge 10^{-2}$ at the scale $\mu^2=4\,{\rm GeV}^2$ for
            the quark flavors $a=u,\,d$ (upper panel) and sea quarks $\bar{u},\,\bar{d},\,s$ (lower panel)
            based on the NLO PDF parametrization \cite{Hou:2019qau}. We show QPDFs for 
            the velocities $v=0.5, \, 0.7,\, 0.9$ and include the PDFs
            for comparison. \newline { }  }   
        \end{minipage}
    \\
    & & \\
      \includegraphics[width=.27\linewidth]{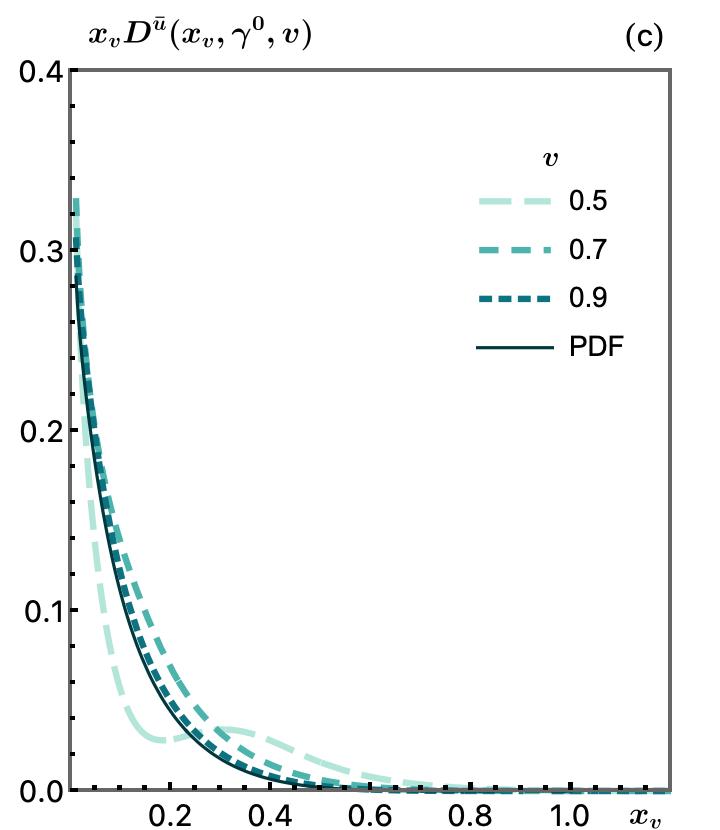}
    & \includegraphics[width=.27\linewidth]{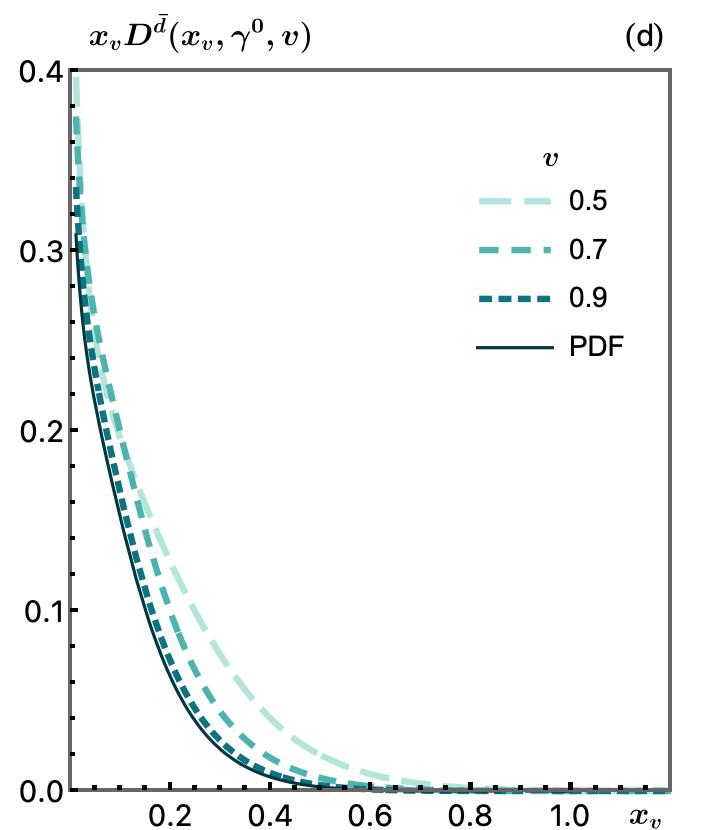} 
    & \includegraphics[width=.27\linewidth]{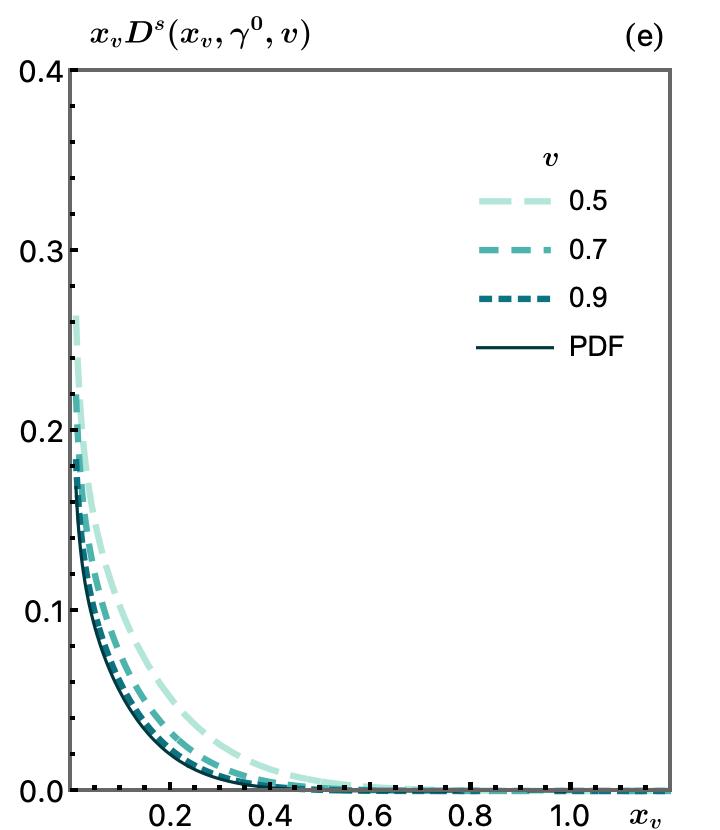}
\end{tabular}
\end{figure}

In Fig.~\ref{Fig-02:QPDF-gamma-0} we show results for $x_vD^a(x_v,\gamma^0,v)$
from the CPM obtained in the above described way. It is of advantage to show
$x_vD^a(x_v,\Gamma,v)$ rather than the QPDFs themselves to suppress their strong 
rise at small $x_v$, and the curves are plotted for positive $x_v \ge 10^{-2}$.
The region of smaller $x_v$ (omitted in Fig.~\ref{Fig-02:QPDF-gamma-0})
will be discussed in detail in Sec.~\ref{Sec:num-small-xv}. 
Notice that for the quark flavors $u$ and $d$ in the upper panel of 
Fig.~\ref{Fig-02:QPDF-gamma-0} we use a different scale than for the 
sea quarks $\bar{u}$, $\bar{d}$, $s=\bar{s}$ in the lower panel of 
Fig.~\ref{Fig-02:QPDF-gamma-0}.

%========== BEGIN FIGURE 2 =======================================
%
%============ END FIGURE 2 =======================================

For the selected velocities, the convergence of the QPDFs toward the corresponding PDFs 
is relatively uniform. The only exception is $x_vD^{\bar{u}}(x_v,\gamma^0,v)$ at the 
velocity $v=0.5$ which shows a ``wiggle'' for $x_v \lesssim 0.3$. This is a ``remnant'' of 
the opposite-sign leaking phenomenon manifest for $x_vD^a(x_v,\gamma^0,v)$ at still lower 
velocities $v \lesssim 0.3$ which was shown in Fig.~\ref{Fig1:leaking-in-toy-model}b 
and discussed in detail in Sec.~\ref{Sec:support-leaking}.
In particular, it is worth noticing that for $v=0.9$ the QPDFs agree with the PDFs for 
$10^{-2} \lesssim x_v \lesssim 0.8$ already within about $5\,\%$ accuracy.
In QCD and other quark models a slower convergence is observed. This indicates the importance
of offshellness effects for the convergence of QPDFs.

%========== BEGIN FIGURE 3 =======================================
%
\begin{figure}[t!]
\begin{tabular}{lll}  
      \includegraphics[width=.27\linewidth]{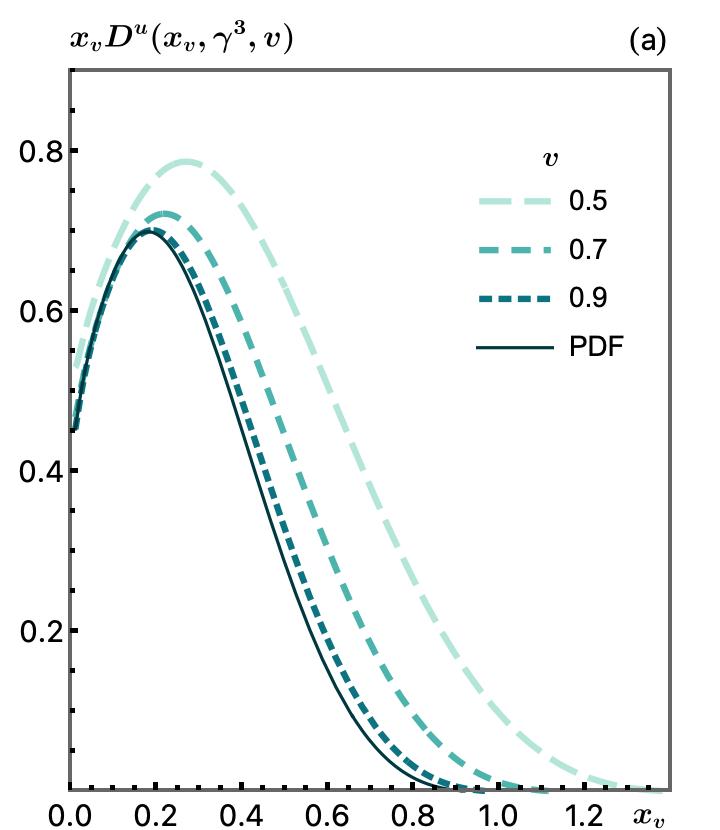}
    & \includegraphics[width=.27\linewidth]{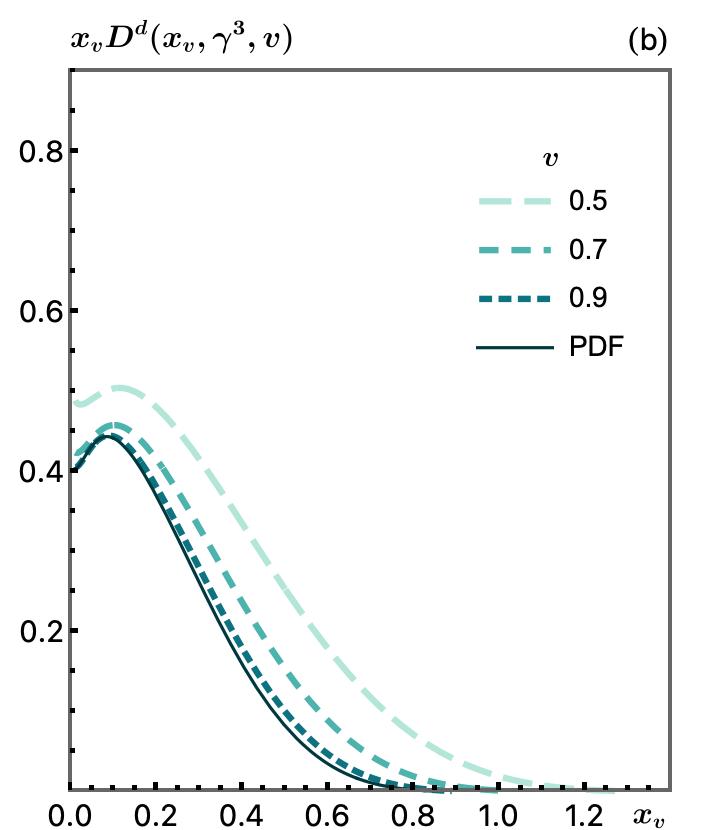}
    &       \begin{minipage}[b]{.32\linewidth}\vspace*{0pt} 
            \caption{\label{Fig-03:QPDF-gamma-3} The QPDFs $x_v D^a(x_v,\gamma^3,v)$ 
            in the CPM as functions of $x_v\ge 10^{-2}$ at the scale $\mu^2=4\,{\rm GeV}^2$ for
            the quark flavors $a=u,\,d$ (upper panel) and sea quarks $\bar{u},\,\bar{d},\,s$ (lower panel)
            based on the NLO PDF parametrization \cite{Hou:2019qau}. We show QPDFs for 
            the velocities $v=0.5, \, 0.7,\, 0.9$ and include the PDFs
            for comparison. \newline { } }
            \end{minipage}
    \\
    & & \\
      \includegraphics[width=.27\linewidth]{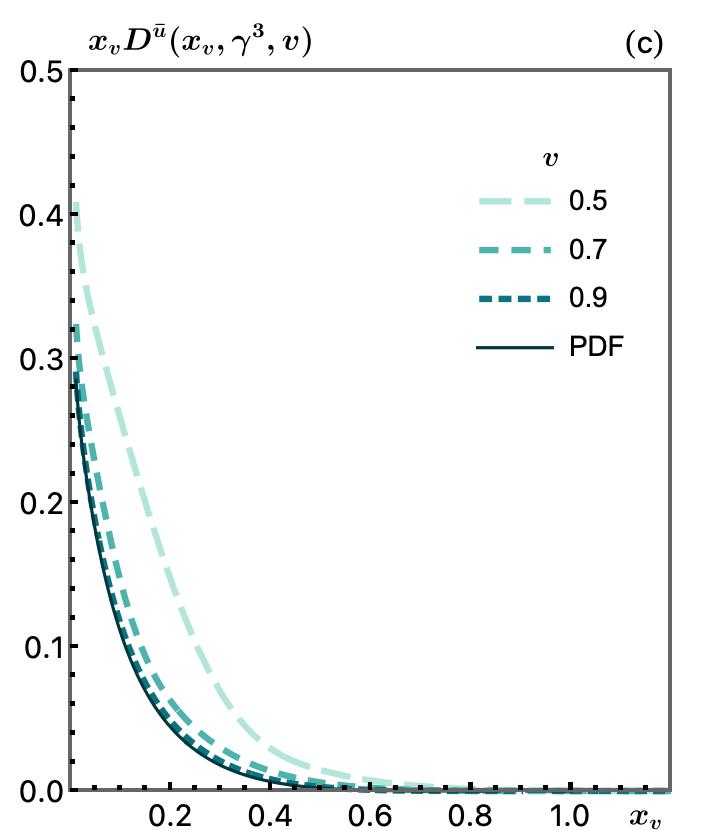}
    & \includegraphics[width=.27\linewidth]{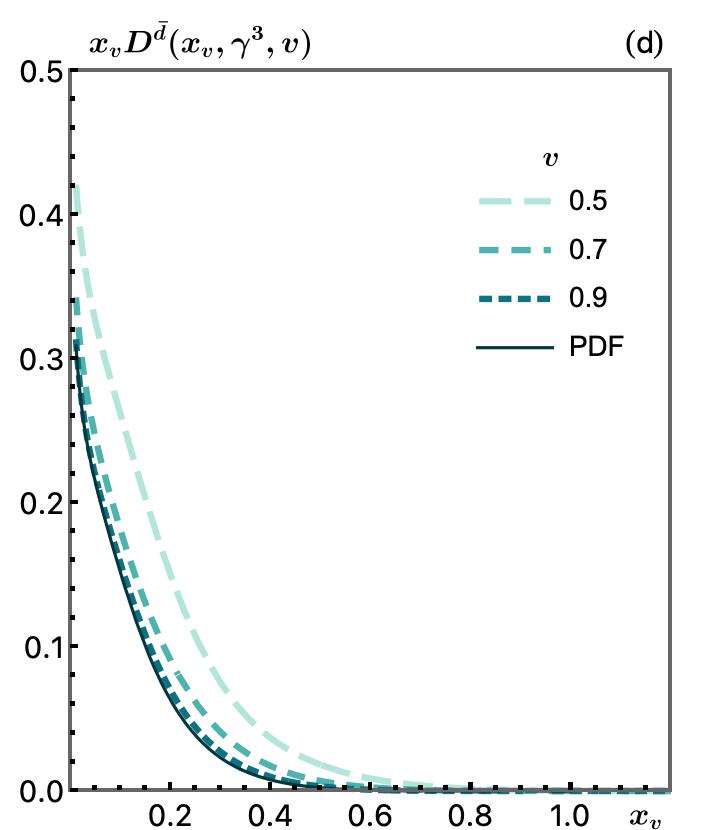} 
    & \includegraphics[width=.27\linewidth]{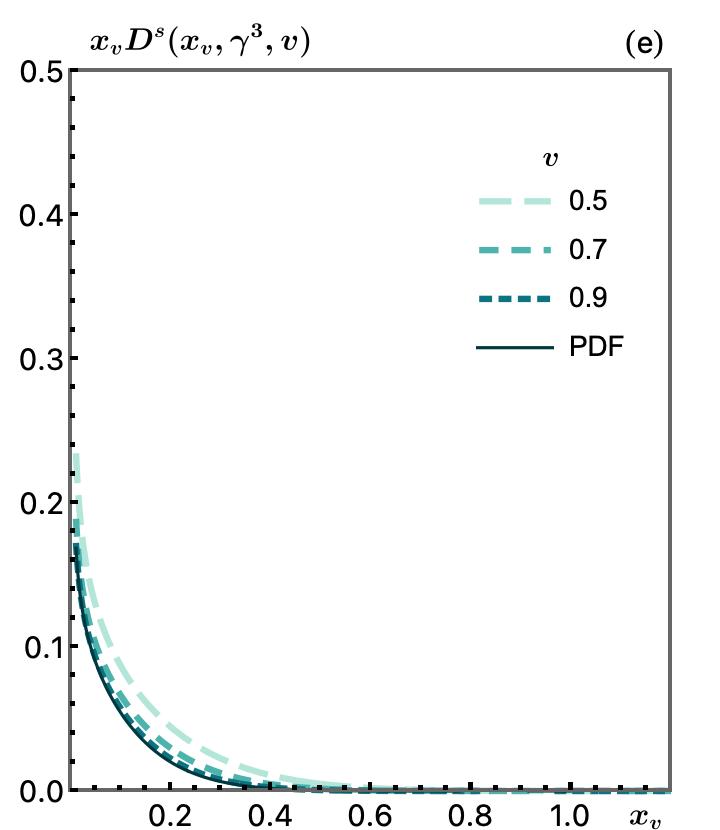}
\end{tabular}
\end{figure}
%============ END FIGURE 3 =======================================

In Fig.~\ref{Fig-03:QPDF-gamma-3} we show the analogous CPM results for $x_vD^a(x_v,\gamma^3,v)$
displaying also in this case only the region of $x_v \ge 10^{-2}$ and postponing
the discussion of the region of smaller $x_v$ for Sec.~\ref{Sec:small-xv}. Notice that also in 
this figure we use different scales for the quark QPDFs in the upper panel and the antiquark QPDFs
in the lower panel of Fig.~\ref{Fig-03:QPDF-gamma-3}. We use the same velocities as in the case 
$\Gamma=\gamma^0$ in Fig.~\ref{Fig-02:QPDF-gamma-0}, namely $v=0.5,\,0.7,\,0.9$ and observe
an even more uniform convergence toward the PDFs as compared to the $\Gamma=\gamma^0$ case.

One interesting observation is that as the limit $v\to 1$ is approached, in
the region $10^{-2}\le x_v \lesssim 0.2$ the quark QPDFs $x_vD^q(x_v,\gamma^3,v)$ for
$q=u,\,d$ converge faster toward the corresponding PDFs than the $x_vD^q(x_v,\gamma^0,v)$. 
Hereby we also notice an interesting flavor-dependence with $x_vD^u(x_v,\gamma^3,v)$ converging 
toward $xf_1^u(x)$ faster than $x_vD^d(x_v,\gamma^3,v)$ converging toward $xf_1^d(x)$, cf.\ 
Figs.~\ref{Fig-03:QPDF-gamma-3}a and \ref{Fig-03:QPDF-gamma-3}b. The latter observation can be
explained by the leaking phenomenon discussed in Sec.~\ref{Sec:support-leaking}. In the
case $\Gamma = \gamma^3$ the leaking occurs with the same sign, but it is important to
recall the sea quark flavor asymmetry $f_1^{\bar d}(x) > f_1^{\bar u}(x)$, while for quarks
we have the opposite situation with $f_1^u(x) > f_1^d(x)$. Therefore, the leaking effect of 
the larger $\bar{d}$-antiquark contribution on the smaller $d$-quark QPDF is more pronounced
than in the $u$-flavor case which causes a slower convergence.
It will be interesting to see if this trend can also be seen in other quark models and lattice
QCD studies. 

It is also noteworthy that we do not observe any wiggles in $x_vD^a(x_v,\gamma^3,v)$ 
which converge uniformly toward the corresponding PDFs as $v$ approaches unity
(to be contrasted with the case for $\Gamma=\gamma^0$ as evident for ${\bar u}$ at $x_v=0.5$).
It is instructive to compare the two cases $\Gamma=\gamma^0$ and $\gamma^3$ more
quantitatively which we shall do in Sec.~\ref{Sec:numerics-compare-gamma-0-vs-3}.

\subsection{Comparison of the \boldmath $\Gamma=\gamma^0$ vs $\gamma^3$ cases}
\label{Sec:numerics-compare-gamma-0-vs-3}

\begin{figure}[b!]
\centering
\begin{tabular}{ccc}
  \includegraphics[width=.27\linewidth]{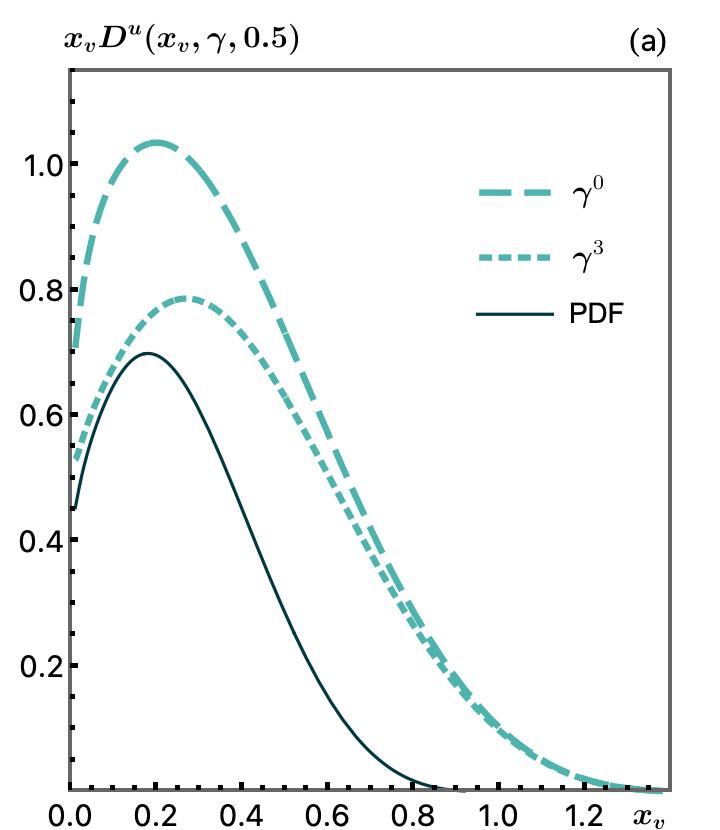} \hspace{0.5cm} &
  \includegraphics[width=.27\linewidth]{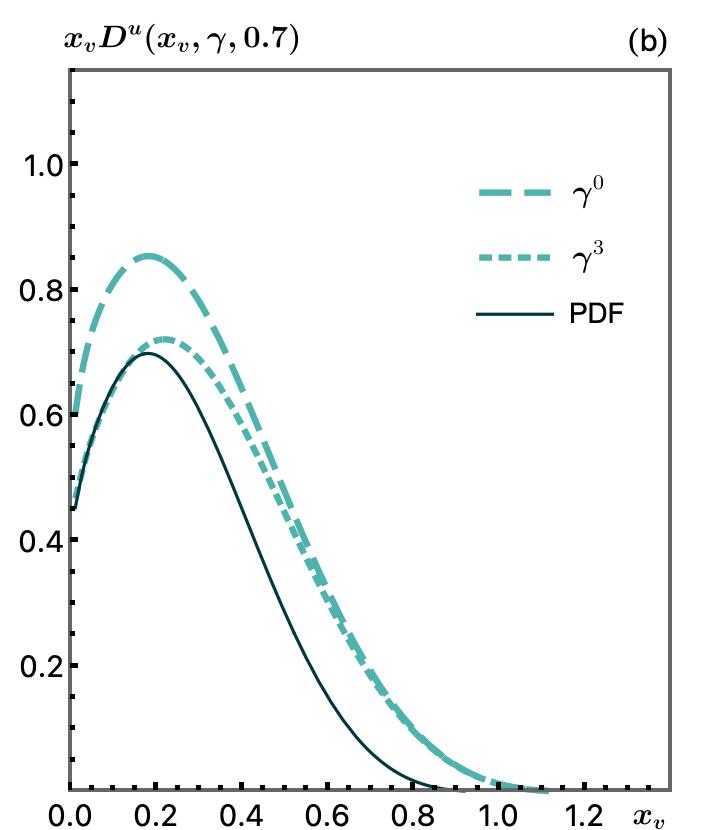} \hspace{0.5cm} & 
  \includegraphics[width=.27\linewidth]{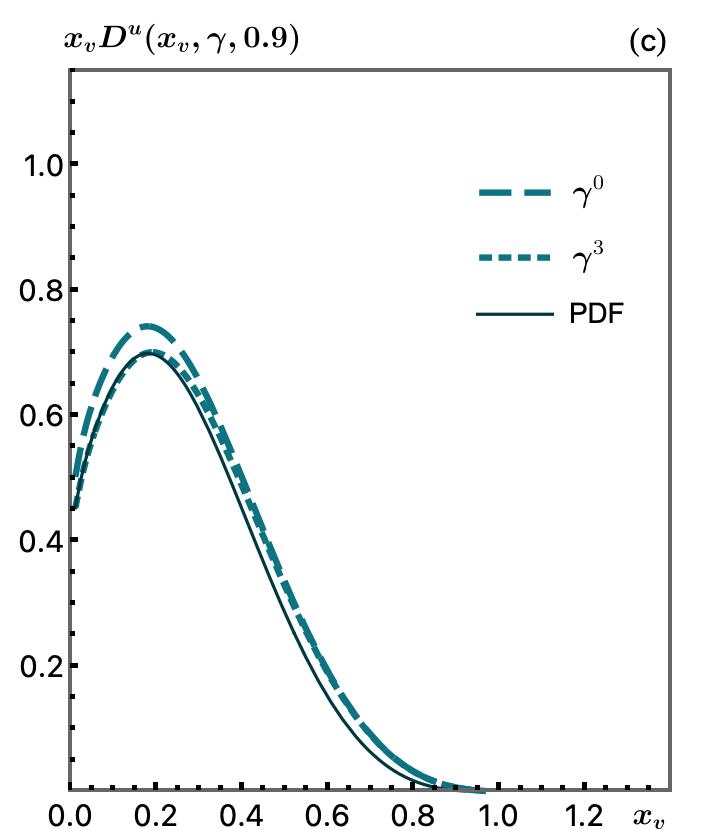} \\
  { } & { } & \\
  \includegraphics[width=.27\linewidth]{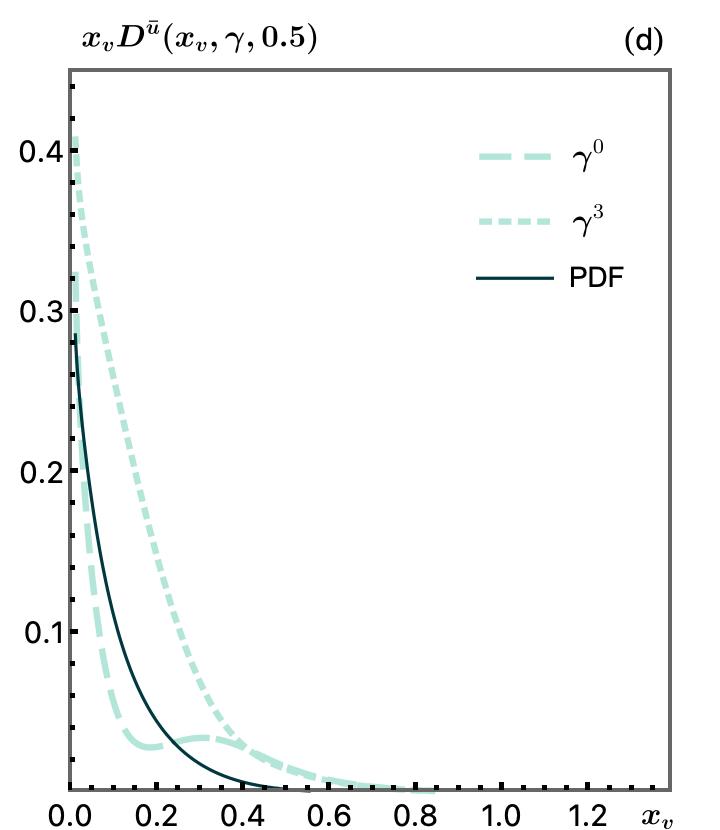} \hspace{0.5cm} &
  \includegraphics[width=.27\linewidth]{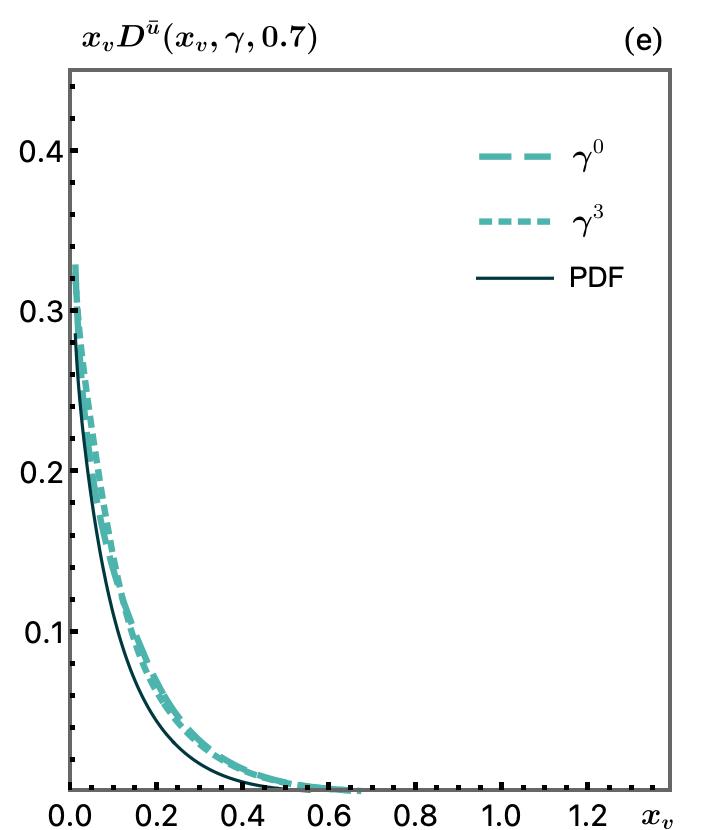} \hspace{0.5cm} &
  \includegraphics[width=.27\linewidth]{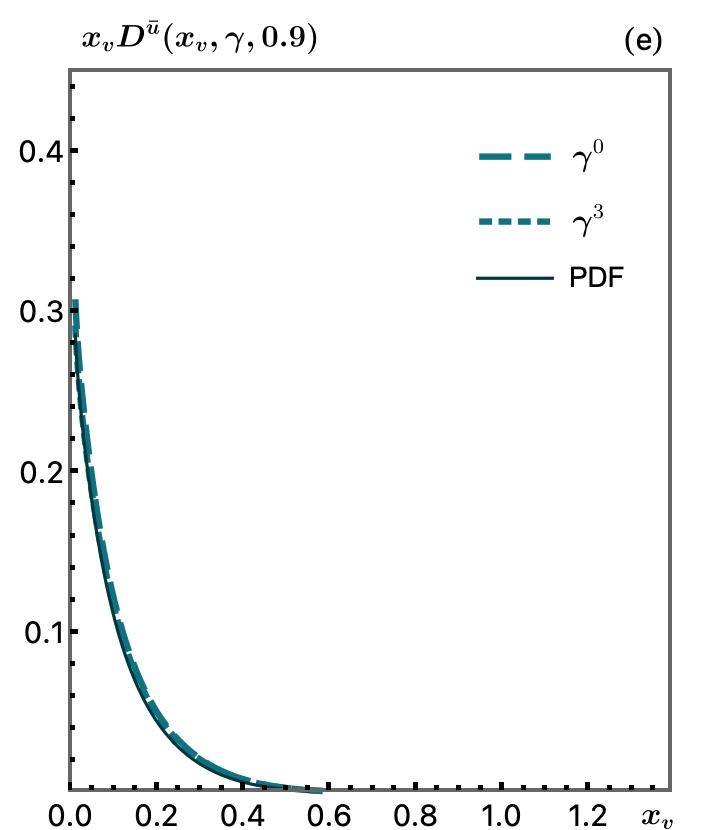}
  \end{tabular}
\caption{\label{Fig4:gamma-comparison}
The QPDFs $x_v D^a(x_v,\Gamma,v)$ for $\Gamma=\gamma^0$ and $\gamma^3$ from the CPM at 
the velocities $v=0.5,\,0.7,\,0.9$ with the PDFs included for comparison.
Upper panel: $u$-quark. Lower panel: $\bar{u}$-antiquark. 
Not shown are the $d$-quark QPDF which looks qualitatively similarly to the $u$-quark case,
and other seaquark QPDFs which exhibit similar patterns to the $\bar{u}$-antiquark case.}
\end{figure}

In Fig.~\ref{Fig4:gamma-comparison} we present a direct comparison of the cases 
$\Gamma=\gamma^0$ and $\gamma^3$ at the selected velocities $v=0.5,\,0.7,\,0.9$
including the respective PDFs for comparison. Hereby, we limit ourselves the 
$u$-quark QPDFs in Figs.~\ref{Fig4:gamma-comparison}a--c and omit the $d$-quark 
QPDFs which look qualitatively very similarly.
We also limit ourselves to showing the $\bar{u}$-antiquark QPDFs
in Figs.~\ref{Fig4:gamma-comparison}d--f representatively for other seaquark 
QPDFs which exhibit similar patterns.

Let us discuss first the quark case. It is an interesting observation that in the
region of larger $x_v$ the two choices $\Gamma=\gamma^0$ and $\gamma^3$ give basically
the same result. What larger $x_v$ means, depends hereby on the velocity. On the scale of
Figs.~\ref{Fig4:gamma-comparison}a--c, for $v=0.5$ the two curves for $\Gamma=\gamma^0$ 
and $\gamma^3$ can hardly be distinguished for $x_v \gtrsim 0.9$. 
When $v=0.7$ this happens for $x_v\gtrsim 0.65$, and for $v=0.9$ the two
curves become hardly distinguishable in the region of $x_v \gtrsim 0.45$. 
Remarkably, while the curves for $\Gamma=\gamma^0$ and $\gamma^3$ agree between
them at larger $x_v$, even at $v=0.9$ they systematically disagree with the PDF. 

%========== BEGIN FIGURE 4 =======================================
%
%
%============ END FIGURE 3 =======================================

Another remarkable observation is that the convergence of $x_vD^u(x_v,\gamma^0,v)$
is relatively uniform over the entire $x_v$ region, see the upper panel of 
Fig.~\ref{Fig4:gamma-comparison}. However, for $x_vD^u(x_v,\gamma^3,v)$ the
situation is distinctly different especially in the smaller-$x_v$ region
which is weakly velocity dependent. For the velocities $0.5\le v \le 0.9$ 
the QPDF $x_vD^u(x_v,\gamma^3,v)$ approaches $xf_1^u(x)$ in the region
$x_v \lesssim (0.2$--$0.3)$ rather rapidly. In fact, on the scale of the
upper panel of Fig.~\ref{Fig4:gamma-comparison} the curves for 
$x_vD^u(x_v,\gamma^3,v)$ and $xf_1^u(x)$ cannot be distinguished for 
$v\ge 0.7$, see  Figs.~\ref{Fig4:gamma-comparison}b and 
\ref{Fig4:gamma-comparison}c.
In particular, the convergence of $x_vD^u(x_v,\gamma^3,v)$ in the 
smaller-$x_v$ region is significantly faster than that of $x_vD^u(x_v,\gamma^0,v)$.
This observation is rather remarkable. 
For the $d$-flavor the situation is very similar. 

Next we discuss the $\Gamma=\gamma^0$ vs $\gamma^3$ comparison for
sea quarks.
Some features are similar to the quark case, but the convergence 
pattern is overall distinctly different for sea quarks. 
Let us begin with the similarity. At the low $v=0.5$ the
$x_vD^{\bar u}(x_v,\gamma^0,v)$ and $x_vD^{\bar u}(x_v,\gamma^3,v)$ 
basically agree with each other for larger $x_v$ (as in the quark case) 
although the onset of this regime with $x_v \gtrsim 0.4$ appears at smaller
$x_v$ compared to the quark case at the same velocity. For $x_v \lesssim 0.4$ 
the antiquark QPDFs differ from each other more significantly than in the quark case. 
The reason for this is the opposite-sign leaking phenomenon in the $\gamma^0$ 
case as compared to the same-sign leaking in the $\gamma^3$ case. The leaking
happens for quark and antiquarks. 
But in the quark case (small $\bar{q}$-effects leak into the large $q$)
the effect is much smaller than in the antiquark case 
(where large $q$-effects leak into the small ${\bar q}$). 
Recall that for still smaller velocities the leaking can even overturn the 
sign of $x_vD^{\bar u}(x_v,\Gamma,v)$, cf.\ the discussion in 
Sec.~\ref{Sec:support-leaking}.

% As $v$ increases, the region of larger-$x_v$ where the $\gamma^0$ and $\gamma^3$ QPDFs extends. 
As $v$ increases, the region of larger-$x_v$ extends where the $\gamma^0$ and $\gamma^3$ QPDFs are numerically close. A remarkable difference between quarks and antiquarks is that the
growth of this region with increasing $v$ proceeds much faster for antiquarks 
such that $x_vD^{\bar u}(x_v,\gamma^0,v)$ and $x_vD^{\bar u}(x_v,\gamma^3,v)$
practically agree with each other (but not with the PDF) already at $v\ge 0.7$.
For still higher velocities the QPDFs (of either choice for $\Gamma$) agree 
with the PDF already within few percent. Thus, the convergence is overall faster 
for antiquark QPDFs than for quark QPDFs in the CPM. 
The case of the $\bar{d}$-flavor is very similar. For the strangeness QPDF,
however, the situation is somewhat different because here the leaking effects 
discriminate far less between the $\gamma^0$ and $\gamma^3$ cases and there
are no sign-turnovers or wiggles at lower velocites, cf.\ the discussion in
footnote~\ref{footnote-sea-type-QPDF}.

\subsection{CPM results in the small \boldmath $x_v$-region}
\label{Sec:num-small-xv}
%========== BEGIN FIGURE 5 =======================================
\begin{wrapfigure}[25]{R}{7cm}
\vspace{-7mm}
\centering
\includegraphics[width=5cm]{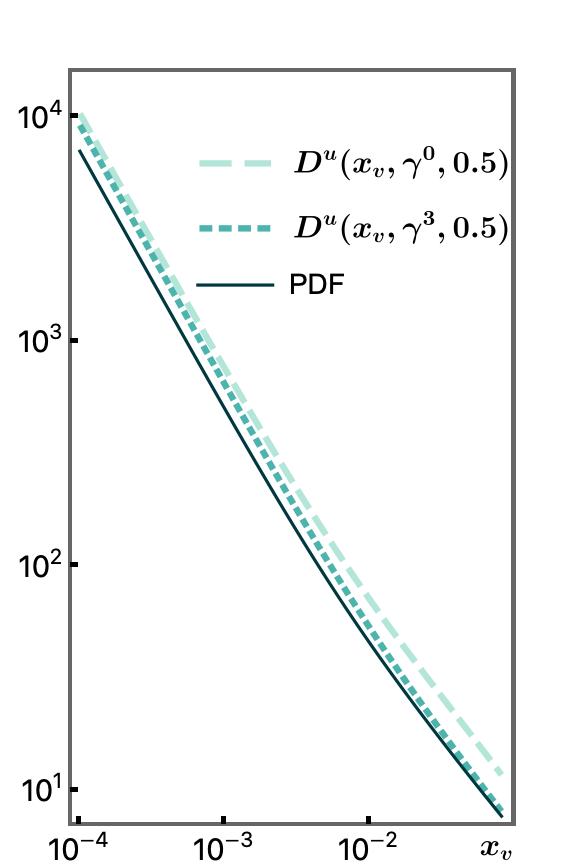}
\caption{\label{Fig-5:small-xv}
$D^u(x_v,\Gamma,v)$ and $f_1^q(x)$ at $v=0.5$ in the small-$x_v$ region. 
The asymptotics in Eqs.~(\ref{Eq:small-x-PDF}--\ref{Eq:small-x-quasi-2}) 
sets in between $10^{-3} < x_v < 10^{-2}$ depending on the function, see text.}
\end{wrapfigure}
%
%============ END FIGURE 5 =======================================

In the CPM, no restrictions arise on the $x_v$-range where QPDFs
can be computed, unlike in other models \cite{Son:2019ghf} or lattice QCD,
see Sec.~\ref{Sec-lattice}. We can therefore apply the model, 
for instance, to the regime of asymptotically small $x_v$.
In Fig.~\ref{Fig-5:small-xv} we show representatively the results for 
$D^u(x,\Gamma,v)$. Other flavors including antiquarks look very similar in the 
small-$x_v$ region. Also $f_1^u(x)$ is shown in 
Fig.~\ref{Fig-5:small-xv}. The pertinent asymptotics from respectively
Eqs.~(\ref{Eq:small-x-PDF},~\ref{Eq:small-x-quasi-1},~\ref{Eq:small-x-quasi-2}) 
are included for comparison. In the small-$x_v$ region, $D^u(x,\gamma^0,v)$
is larger than $D^u(x,\gamma^3,v)$ for velocities $v\gtrsim 0.5$. 
For lower velocities it is vice versa.

The asymptotics of $D^u(x,\gamma^0,v)$ sets in at low $x_v\lesssim 10^{-3}$, 
that of the PDF at $x\lesssim 3\times 10^{-3}$, while for $D^u(x,\gamma^3,v)$ 
this happes at $x_v\lesssim 10^{-2}$. The curves for $D^u(x,\gamma^0,v)$
and $D^u(x,\gamma^3,v)$ at $v=0.5$ are closer to each other than to the PDF. 
The trend that $D^u(x,\gamma^3,v)$ converges faster than $D^u(x,\gamma^0,v)$
toward the PDF in the region $10^{-2} < x_v \lesssim 0.2$ observed in 
Fig.~\ref{Fig4:gamma-comparison}a (and clearly visible also in
Fig.~\ref{Fig-5:small-xv} for $x_v>10^{-2}$)
does not continue into the region of very small-$x_v$.
The results for the $\bar{u}$ QPDFs have exactly the same asymptotics, as otherwise the
flavor number sum rule would diverge, but the asymptotics set in significantly earlier
due to the absence of a valence-$x_v$ region. For $d$ and $\bar{d}$ the results are
very similar with somewhat different numerical values for the constants $A$ and $B$ in
Eqs.~(\ref{Eq:small-x-PDF},~\ref{Eq:small-x-quasi-1},~\ref{Eq:small-x-quasi-2}).
The $s$ and $\bar{s}$ results are analogous to the other sea quark results.

\subsection{Test of Radyushkin formula in Eq.~(\ref{Eq:Radyushkin-formula})}
%========== BEGIN FIGURE 6 =======================================
%
\begin{wrapfigure}[16]{r}{6.5cm}
\vspace{-8mm}
\centering
    \includegraphics[width=5cm]{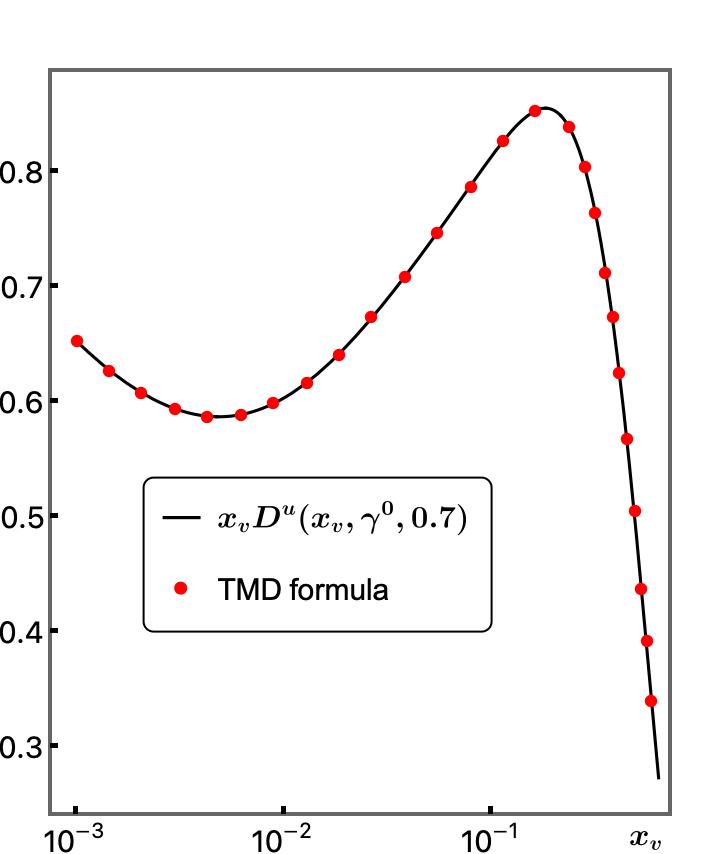}
    \caption{$x_v\,D^u(x_v,\gamma^0,v)$ at $v=0.7$ in the CPM
    computed directly (solid line) and via Eq.~(\ref{Eq:Radyushkin-formula})
    (discrete points). \label{Fig-Radyushkin}}
\end{wrapfigure}
%============ END FIGURE 6 =======================================

The relation (\ref{Eq:Radyushkin-formula}) provides an important cross check 
for the numerics and serves as a theoretical consistency test of the model.
The model expression for the unpolarized quark TMD is given by
\cite{Efremov:2009ze}
\ba
    f_1^q(x,\vec{k}_\perp^{\,2}) 
%    &=& xM \int \frac{dk^3}{|\vec{k}^{\,}|}\,
%    \Theta(M-2|\vec{k}^{\,}|)\,{\cal G}^q(M|\vec{k}^{\,}|)\,
%    \delta\biggl(x-\frac{|\vec{k}^{\,}|+k^3}{M}\biggr)\nonumber\\
    &=& xM^2\,\Theta\Bigl(x(1-x)M^2-\vec{k}_\perp^{\,2}\Bigr)\,
    {\cal G}^q\biggl(\frac{xM^2}{2}+\frac{\vec{k}^{\,2}_\perp}{2x}\biggr)
\ea
where the argument of ${\cal G}^q$ is $P\cdot k = Mk^0$ with 
$k^0=\frac{xM}{2}(1+\frac{\vec{k}_\perp^{\,2}}{x^2M^2})$.

Computing a QPDF via the Radyushkin relation is numerically more 
involved, since for every $x_v$-value one needs to determine the TMD at all 
values of $x$ and $\vec{k}_\perp^{\,2}$ and then carry out integrations over 
$dx$ and $d^2k_\perp$. Even in simple models, this takes up significantly more 
computing time than the direct calculation. We therefore content ourselves with
computing the QPDF at only selected $x_v$ shown as bullets in 
Fig.~\ref{Fig-Radyushkin}. The Radyushkin formula is valid in the CPM
within numerical accuracy. 

Note that in Sec.~\ref{Sec:Radyushkin-formula} we have analytically proven 
that this relation holds in any quark model including the CPM.

\subsection{Comparison to lattice QCD results}
\label{Sec-lattice}

\begin{figure}[t!]
\centering
\includegraphics[width=8cm]{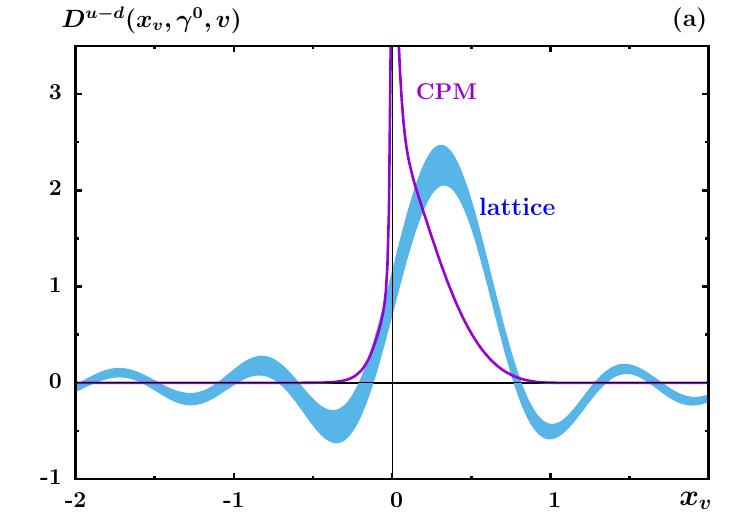}
\includegraphics[width=8cm]{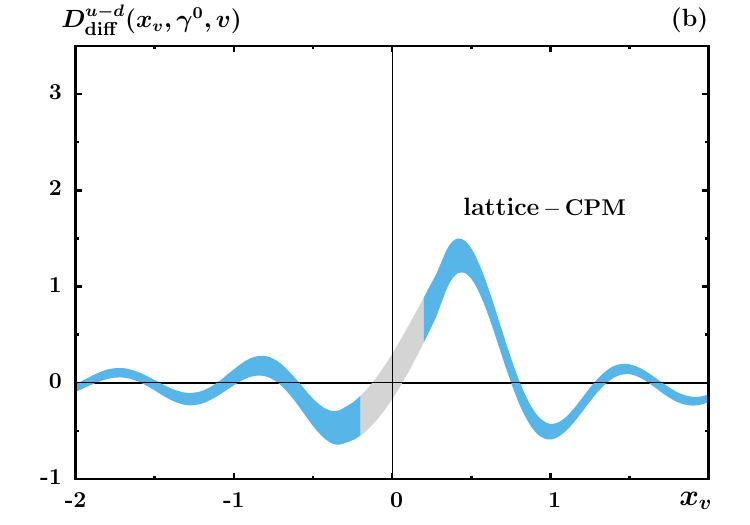}
\caption{(a) Unpolarized isovector QPDF $(D^u-D^d)(x_v,\gamma^0,v)$ 
at a scale of $\mu^2=4\,{\rm GeV}^2$ as function of $x_v$ for 
$P_z=1.38\rm GeV$ which corresponds $v = 0.827\,c$ from the lattice 
QCD study of Ref.~\cite{Alexandrou:2019lfo} in comparison to the 
CPM results obtained in this work.
(b) Difference of lattice result minus model result.
The light-shaded region $|x|<0.3$ is filled in to guide the eye, see text.
\label{Fig-lattice}}
\end{figure}

Finally, it is instructive to present a comparison to a lattice QCD calculation. 
In Fig.~\ref{Fig-lattice}a we show the model result for the unpolarized isovector 
QPDF $(D^u-D^d)(x_v,\gamma^0,v)$ at a scale of $\mu^2=4\,{\rm GeV}^2$ in comparison
to the lattice QCD study \cite{Alexandrou:2019lfo} at the largest available value 
of nucleon momentum in \cite{Alexandrou:2019lfo}, $P_z=1.38\,$GeV, corresponding 
to a nucleon velocity of $v = 0.827\,c$. This lattice calculation was performed on 
a $48^3\times 96$ lattice with a lattice spacing $a= 0.094\,$fm, a lattice size of 
$L = 4.5\,$fm, and a practically physical pion mass of 130$\,$MeV. 

Our simple model cannot be expected to agree exactly with lattice QCD. 
However, the model QPDFs converge by default into the physical PDFs, 
and the lattice QCD results are ultimately expected to do the same. 
Our covariant model correctly describes the relativistic ``kinematics'' 
but it misses the ``dynamics'' owing to the onshell assumption.  
It is therefore instructive to investigate the difference between the lattice
and model results in the hope to learn something about the size of offshellness
effects. We do this in Fig.~\ref{Fig-lattice}b where we plot 
\be
    D^{u-d}_{\rm diff}(x_v,\gamma^0,v) = 
    (D^u-D^d)(x_v,\gamma^0,v)|_{\rm lattice}-
    (D^u-D^d)(x_v,\gamma^0,v)|_{\rm CPM} 
\ee
with an uncertainty band due to the lattice data \cite{Alexandrou:2019lfo}. 
The light-shaded region in the interval $|x_v|<0.3$ indicates the region where
the lattice calculation is likely not reliable as explained below, and we 
removed this part interpolating with cubic splines to guide the eye. 
With this reservation in mind, Fig.~\ref{Fig-lattice}b seems to suggest that 
we may see a superposition of two effects, one being undoubtedly due to the model 
assumption of free partons and another possibly due to systematic effects in  
\cite{Alexandrou:2019lfo}. Despite great efforts to control such effects, 
lattice calculations of QPDFs are still at a pioneering stage and the presence of 
uncontrolled systematic uncertainties cannot be excluded. The truncation of the 
Fourier transform is among the most challenging systematic effects 
\cite{Alexandrou:2019lfo} as discussed below.

The QPDF was computed in \cite{Alexandrou:2019lfo} by evaluating the matrix element
$h^q(\nu,\gamma^0,P_z)=\frac12\langle N_v|\,\overline{\Psi}^{\,q}(0){\cal W}\gamma^0\;\Psi^q(z)\,|N_v\rangle$ for $z^\mu=(0,0,0,z^3)$ with $\nu = P^3 z^3$ which yields the QPDF via
\be\label{Eq:def-Fourier}
    D^q(x_v,\Gamma,v) = \int\frac{d\nu}{2\pi}\;e^{-i\nu x_v^{ }}\;h^q(\nu,\Gamma,P_z)\,.
\ee
This corresponds to the quark model expression in Eq.~(\ref{Eq:def-quasi-I})
except for the appearance of the Wilson line ${\cal W}$.
More precisely, in lattice calculations the Fourier transform is evaluated as a discrete 
series rather than a continuous integral. More importantly, due to the finite lattice size,
the integration (or discrete summation) goes over a finite interval 
$-\nu_{\rm max} < \nu < \nu_{\rm max}$ with $\nu_{\rm max} = P_z z_{\rm max}=6.5$ for 
$P_z=10\frac{\pi}{L}=1.38\,\rm GeV$ and $z_{\rm max}=10\,a$ in Ref.~\cite{Alexandrou:2019lfo}.
Since large $|\nu|$ are correlated with small $|x_v|$, the truncation at a finite $\nu_{\rm max}$ 
implies limitations to reach the small-$x_v$ region of QPDFs on a finite lattice.\footnote{%
    This is independent of the limitation when determining PDFs from QPDFs in the 
    limit $v\to 1$ or $P_z\to\infty$ in a matching procedure in which power corrections 
    of the type $M^2/(x^2(1-x)P_z^2)$ emerge which are particularly large for $x\to 0$ and $x\to 1$     \cite{Lin:2017snn,Monahan:2018euv,Cichy:2018mum,Zhao:2018fyu,Constantinou:2020pek,Constantinou:2022yye}.} 
Two methods were used in \cite{Alexandrou:2019lfo} to evaluate the Fourier 
transform, the ``standard'' and ``derivative'' method of Ref.~\cite{Lin:2017ani}.
The QPDFs obtained with the two methods agree for $|x_v|\gtrsim 0.3$, see Fig.~30 in 
\cite{Alexandrou:2019lfo}, which we used for our rough estimate of the ``less reliable region''
light-shaded $|x_v|$-region in our Fig.~\ref{Fig-lattice}b. The result from \cite{Alexandrou:2019lfo} 
shown in our Fig.~\ref{Fig-lattice}b was obtained using the standard method of Ref.~\cite{Lin:2017ani}.

The oscillatory behavior in Fig.~\ref{Fig-lattice}b might be related to truncating 
the Fourier transform. Other methods to address this issue can be pursued, such as 
the Backus-Gilbert method or neural-network based methods, see \cite{Chu:2025jsi} 
and references therein for the latest developments. 
Further dedicated studies and inclusion of higher $P_z$ values may be needed to shed light
whether the oscillatory  behavior in Fig.~\ref{Fig-lattice}b is due to QCD offshellness effects 
or whether also systematic Fourier transform truncation effects may contribute. 
The unambiguous determination of QCD offshellness effects would provide valuable guidance
how to relax the onshell condition in the CPM and potentially obtain a more realistic model.

\newpage
\section{Conclusions}

We presented a study of QPDFs in quark models, defined as models 
without explicit gauge degrees of freedom. We assumed that the models 
(i) are Lorentz invariant, and (ii) the model expressions for the $A_i(P\cdot k,k^2)$
amplitudes describing the quark correlator are either finite or made finite by means 
of appropriate regularization or renormalization such that Lorentz invariance 
is preserved. Otherwise we made no assumptions whatsoever
about the model dynamics. We have shown that in all such models the unpolarized QPDFs 
are consistently described, yield PDFs when the nucleon velocity $v\to1$ is taken, 
satisfy quark flavor and momentum sum rules (if the PDFs do so in a given model),
and obey the Radyushkin relation. 
One interesting observation was that the QPDFs in such models are elegantly and 
efficiently described using the nucleon velocity $v$ rather than the nucleon momentum 
$P_z$ as a variable. Another interesting observation is that the QPDF with $\Gamma=\gamma^3$ 
contains fewer explicit power corrections than that with $\Gamma=\gamma^0$.

We have then chosen the covariant parton model (CPM) as an illustration
which takes PDFs (from parametrizations) as input and allows one to compute 
QPDFs of quarks and antiquarks under the assumption that they are non-interacting 
partons, i.e.\ in a Wandzura-Wilczek (type) approximation. Benefiting from the 
simplicity and lucidity of this model, we have made a number of interesting observations. 
For instance, while the quark and antiquark effects are clearly separated in PDFs, they 
contaminate each other in the case of QPDFs. We presented numerical results for the QPDFs 
of $u$, $d$, $\bar u$, $\bar d$, $s$, $\bar s$ flavors and investigated their convergence. 
In the region of larger $x_v\gtrsim 0.3$ 
% at $\mu^2=4\,\rm GeV^2$ 
the two choices $\Gamma=\gamma^0$
or $\Gamma=\gamma^3$ are very similar. In the region $x_v\lesssim 0.3$ the QPDF with 
$\Gamma=\gamma^3$ converges toward the PDF significanty faster. 
We also studied the small-$x_v$ behavior which can be done analytically in this model but 
is out of reach in lattice QCD. As an interesting byproduct we have shown that the 
energy-momentum tensor form factor $\bar{c}^q(t)$ satisfies $\bar{c}^q(0)=-\tfrac14\,A^q(0)$
in the CPM and hence also in the Wandzura-Wilczek approximation.
The CPM results entail the complete target-mass corrections.

We also presented a comparison to a lattice QCD calculation. Since the model results 
for QPDFs converge to the correct PDFs as do the lattice results, we investigated the 
difference of lattice and model results in the hope to obtain insight on the size
of offshellness effects in QPDFs. At the present stage, it is unclear whether the 
observed difference is due to offshellness effect or might include also systematic 
effects which are not yet under full control in pioneering lattice studies like,  
e.g., effects due to the truncation of the Fourier transform. Insights on offshellness 
effects would be very interesting as they could guide novel developments to relax 
the onshell restriction in the CPM and construct more realistic models. 

The advantage of the CPM model framework is its fully field-theoretical description of
the unintegrated quark correlator which allows one to describe QPDFs and TMDs in terms
of PDFs which serve as input for this model. The next steps will include applications to 
polarized QPDFs and subleading twist. The results might give helpful insights for the
extrapolation of QPDFs in lattice QCD.

\ \\
{\bf Acknowledgments.} 
We thank Shohini Bhattacharya, Krzysztof Cichy, and Martha Constantinou for 
valuable discussions and making the lattice results of \cite{Alexandrou:2019lfo}
available. This work was supported by the NSF Award No.\ 2412625 and the DOE  
Quark-Gluon Tomography Topical Collaboration with Award No.\ DE-SC0023646.

\appendix

\section{Conventions of QPDF\lowercase{s} in literature}
\label{App:conventions}

Different conventions have been used in literature. For clarity and to reduce 
confusion, we provide some comments and comparisons here. In the velocity
formulation of QPDFs, the different conventions can be simply stated as
\be\label{App:conv-1}
      D^q(x_v,\Gamma,v)\Bigl|_{\mbox{\footnotesize this work}} 
    = D^q(x_v,\Gamma,v)\Bigl|_{\mbox{\footnotesize Ref.~\cite{Ji:2013dva}}} 
    = \frac1v\,D^q(x_v,\Gamma,v)\Bigl|_{\mbox{\footnotesize Ref.~\cite{Diakonov:1997vc,Son:2019ghf}}} \,. 
\ee
In Refs.~\cite{Radyushkin:2016hsy,Radyushkin:2017cyf} only the QPDF for the case
$\Gamma=\gamma^0$ was discussed. The connection to our notation is as follows
\be\label{App:conv-2}
     D^q(x_v,\gamma^0,v)\Bigl|_{\mbox{\footnotesize this work}} 
    = \frac1v\,D^q(x_v,\gamma^0,v)\Bigl|_{\mbox{\footnotesize Refs.~\cite{Radyushkin:2016hsy,Radyushkin:2017cyf}}} \,. 
\ee
One convenient way to quickly verify the underlying conventions, is to 
compare the sum rules $\int dx_v^k D^q(x_v,\Gamma,v)$ for $k=0,\;1$ in 
Eqs.~(\ref{Eq:general-mom1-quasi-0},~\ref{Eq:general-mom1-quasi-3},~\ref{Eq:general-mom2-quasi-0},~\ref{Eq:general-mom2-quasi-3})
in our work vs the sum rules in Eqs.~(21,~25) of \cite{Son:2019ghf}
or Eqs.~(2.25,~2.28) of \cite{Radyushkin:2016hsy} which differ by factors
of $v$ according to Eqs.~(\ref{App:conv-1},~\ref{App:conv-2}).

As a consequence of different conventions, the formula (\ref{Eq:Radyushkin-formula}) 
relating QPDFs and TMDs is formulated in our convetion with the prefactor $P_0$ while in 
\cite{Radyushkin:2016hsy} it was originally derived with the prefactor of $P_z = vP_0$.

\newpage

\section{Dirac form factor \boldmath $F_1^q(0)$}
\label{App:em-FF}

In the Appendices, the four-momenta $P^\mu$ and $k^\mu$ are understood 
in a general reference frame unless we explicitly specify otherwise. 
In this Appendix  we derive the expression for the Dirac form factor
in the forward limit. 

The contribution of the quark flavor $q$ 
to the electromagnetic current operator is given by 
$J^\mu_q(y) = \overline{\Psi}_q(y)\gamma^\mu\Psi_q(y)$. 
The nucleon matrix elements of this operator are described in 
terms of the electromagnetic form factors as follows
\be 
    \la N^\prime|J_q^\mu(y)|N\rangle
    = \bar u_{N'}\biggl[\gamma^\mu F_1^q(t)
    + \frac{i \sigma_{\mu\nu}\Delta_\nu}{2M}\, F_2^q(t)\biggr]u_N\,
    e^{i\Delta\cdot y},
    \label{Eq:ff-of-em}
\ee
where $\Delta^\mu=P^{\prime \mu}-P^\mu$ and $t=\Delta^2$
and $|N\ra=|P,S\rangle$ denotes the nucleon state. The nucleon 
spinors $u_N=u(P,S)$ are normalized as $\bar u_{N}\,u_N=2M$. 
Using the identity $\bar u_N\,\gamma^\mu \,u_N=2P^\mu$ we obtain
in the forward limit the expression
\be \label{Eq:ff-of-em-2}
    \la N|J_q^\mu(y)|N\rangle = 2P^\mu \,F_1^q(0) \,
\ee
which we wish to evaluate in quark models in terms of the 
$A_i^q(P\cdot k,k^2)$ amplitudes. Notice that translation invariance 
allows us to arbitrarily shift the position $y$ of the operator which introduces an overall phase in off-forward case  (\ref{Eq:ff-of-em}), 
and literally has no effect in forward case (\ref{Eq:ff-of-em-2}). 
It is convenient to choose $y=0$. 

In order to derive the expressions for $\la N|J_q^\mu(y)|N\rangle$
in quark models in terms of amplitudes $A_i^q$, we proceed as
follows. Based on Eq.~(\ref{Eq:correlator-q}) we obtain 
\be\label{Eq:ff-of-em-3}
    \int \frac{d^4z}{(2\pi)^4}\;e^{i k\cdot z}\,
    \langle N|\,\overline{\Psi}^{\,q}(0)\,\gamma^\mu\,\Psi^q(z)\,
    |N\rangle
    = {\rm tr}\bigl[\Phi^q(k,P,S)\,\gamma^\mu\bigr]\,.
\ee
The Fourier transform of this expression yields
\be\label{Eq:ff-of-em-4}
    \langle N|\,\overline{\Psi}^{\,q}(0)\,\gamma^\mu\,\Psi^q(z)\,
    |N\rangle
    = \int d^4k\;e^{-i k\cdot z}\,
    {\rm tr}\bigl[\Phi^q(k,P,S)\,\gamma^\mu\bigr]\,.
\ee
Being interested in the local operator
$J^\mu_q(0) = \overline{\Psi}_q(0)\gamma^\mu\Psi_q(0)$
we set  $z=0$ in Eq.~(\ref{Eq:ff-of-em-4}) and obtain
\be\label{Eq:ff-of-em-5}
    \langle N|\,J_q^\mu(0)\,|N\rangle
    = 2P^\mu \,F_1^q(0)
    = \int d^4k\;{\rm tr}\bigl[\Phi^q(k,P,S)\,\gamma^\mu\bigr]
    = 4 \int d^4k\;\biggl(P^\mu\,A_2^q + k^\mu\,A_3^q\biggr) \,,
\ee
where in the last step we made use of Eq.~(\ref{Eq:tr-phi-gamma-mu}).
We now choose to evaluate Eq.~(\ref{Eq:ff-of-em-5}) in the nucleon 
rest frame specified by Eq.~(\ref{Eq:P-k-rest-frame}). This yields
\be\label{Eq:ff-of-em-6}
    F_1^q(0)
    = 2 \int d^4k\;\biggl(A_2^q + \frac{k^0}{M}\,A_3^q\biggr) \,.
\ee
Notice that the expression in Eq.~(\ref{Eq:ff-of-em-5}) contains 
under the $d^4k$ integral terms proportional to $k^i A_3^q$
which are odd functions and drop out after 
integration. Only the expectation value of the $J^0_q(0)$ 
component yields a non-zero result which corresponds to the 
electric charge. It is customary to define 
\be\label{Eq:ff-of-em-7}
    F_1^q(0) =  N^q
\ee
with $N^u=2$ and $N^d=1$ for proton (vice versa for neutron).
The convention is that the nucleon Dirac form factor is given by 
$F_1(t)=\sum_q e_q \,F_1^q(t)$ where $e_u=\frac23$, etc are the
quark fractional charges. At $t=0$ one recovers the electric charge 
of the particle in units of the elementary unit of charge, i.e.\ 
the value $F_1(0)=1$ for proton (zero for neutron). 

We notice that that the expression for $F_1^q(0)=N^q$ was encountered 
in the sum rules in Eqs.~(\ref{Eq:mom1-PDF}-\ref{Eq:mom1-quasi-3})
which completes the proof of the quark flavor sum rules in
Eqs.~(\ref{Eq:general-mom1-PDF},~\ref{Eq:general-mom1-quasi-0},~\ref{Eq:general-mom1-quasi-3}).

\section{Energy-momentum tensor form factors \boldmath $A^q(0)$ and $\bar{c}^q(0)$}
\label{App:EMT-FF}

With the notation introduced in the previous section and the definition
$\bar{P}^\mu=\frac12(P^{\prime\mu}+P^\mu)$ the nucleon form factors 
of the quark contribution to the energy-momentum tensor, for a review 
see e.g.\ \cite{Polyakov:2018zvc,Burkert:2023wzr}, can be defined as
\be
    \la N^\prime| \hat T_q^{\mu\nu}(0) |N\rangle
    = \bar u_{N'}\biggl[A^q(t)\,\frac{\bar{P}_\mu \bar{P}_\nu}{M}+
    J^q(t)\ \frac{i(\bar{P}_{\mu}\sigma_{\nu\rho}+\bar{P}_{\nu}\sigma_{\mu\rho})
    \Delta^\rho}{2M}
    + D^q(t)\,
    \frac{\Delta_\mu\Delta_\nu-g_{\mu\nu}\Delta^2}{4M}+
    \bar{c}^q(t)\,M\,g^{\mu\nu}\biggr]u_N\, .
    \label{Eq:ff-of-EMT}
\ee
We are interested in the forward limit of this expression 
which is given by
\be
    \la N| \hat T_q^{\mu\nu}(0) |N\ra
    = 2\,P_\mu P_\nu\,A^q(0)
    + 2M^2 \, g^{\mu\nu}\,\bar{c}^q(0)\, .
    \label{Eq:ff-of-EMT-2}
\ee
Notice that the zero on the left-hand side is the position of
the operator $T^{\mu\nu}_q(y)$ which based on translational invariance
can be conveniently chosen $y=0$, while the 
zero on the right-hand side is the momentum transfer square $t=0$.

While the total energy-momentum tensor operator depends on the details
and degrees of freedom in a model, the expression for the quark part 
of this operator is given in all quark models by 
\be\label{Eq:EMT-symmetric-q}
    T^{\mu\nu}_q(0) = \frac{1}{4}\,
    \overline{\Psi}_q(0)\biggl(
    -i\overset{ \leftarrow}\partial{ }^\mu\gamma^\nu
    -i\overset{ \leftarrow}\partial{ }^\nu\gamma^\mu
    +i\overset{\rightarrow}\partial{ }^\mu\gamma^\nu
    +i\overset{\rightarrow}\partial{ }^\nu\gamma^\mu
    \biggr)\Psi_q(0)\,,
\ee
where the arrows indicate whether $\overline{\Psi}_q(0)$ or $\Psi_q(0)$
are to be differentiated. In order to evaluate the matrix elements of this
operator we carry out the following steps
\ba \label{Eq:correlator-T-mu-nu-1}
    \int \frac{\mathrm{d}^4z}{(2\pi)^4}\;\mathrm{e}^{i k\cdot z}\,
    \langle N|\,\overline{\Psi}^{\,q}(0)\;\gamma^\mu
    \bigl(i\overset{\rightarrow}\partial{ }^\nu\Psi^q(z)\bigr)\,
    |N\rangle
    = {\rm tr}\biggl[\Phi^q(k,P,S)\gamma^\mu k^\nu\biggr] \,,\\
    \label{Eq:correlator-T-mu-nu-2}
    \int \frac{\mathrm{d}^4z}{(2\pi)^4}\;\mathrm{e}^{-ik\cdot z}\,
    \langle N|\,
    \bigl(-\overline{\Psi}^{\,q}(z)\,i\overset{\leftarrow}\partial{ }^\nu)
    \;\gamma^\mu\,\Psi^q(0)\,
    |N\rangle
    = {\rm tr}\biggl[\Phi^q(k,P,S)\gamma^\mu k^\nu\biggr]\,.
\ea
In Eq.~(\ref{Eq:correlator-T-mu-nu-1}) on the left-hand side we perform
an integration by parts, and use the definition of the correlator 
(\ref{Eq:correlator-q}).
In Eq.~(\ref{Eq:correlator-T-mu-nu-2}) we perform an integration by parts, 
use translation invariance to shift the positions of the operator as 
$\langle N|\,\overline{\Psi}^{\,q}(z)\,\gamma^\mu\,\Psi^q(0)|N\rangle=
\langle N|\,\overline{\Psi}^{\,q}(0)\,\gamma^\mu\,\Psi^q(-z)|N\rangle$
and substitute $z^\mu\to(-z^\mu)$ before using the correlator definition.
Inverting the Fourier transforms in 
Eqs.~(\ref{Eq:correlator-T-mu-nu-1},~\ref{Eq:correlator-T-mu-nu-2}),
repeating the steps with exchanged indices $\mu\leftrightarrow\nu$, 
adding up all pieces and taking finally the limit $z^\mu \to 0$ to 
recover the local operator $T_q^{\mu\nu}(0)$ we obtain from
Eq.~(\ref{Eq:ff-of-EMT-2})
\be\label{Eq:correlator-T-mu-nu-3}
    \la N| \hat T_q^{\mu\nu}(0) |N\rangle
    = 2\,P_\mu P_\nu\,A^q(0) + 2M^2\,g^{\mu\nu}\,\bar{c}^q(0)
    = 4 \int d^4k\;
    \biggl(\frac{P^\mu k^\nu+P^\nu k^\mu}{2} A_2^q + k^\mu k^\nu A_3^q\biggr)\,.
\ee
In order to derive the expressions for the form factors $A^q(0)$ 
and $\bar{c}^q(0)$ we contract Eq.~(\ref{Eq:correlator-T-mu-nu-3}) 
with respectively $g^{\mu\nu}$ and $P^\mu P^\nu$ which yields two
equations for two unknowns
\begin{alignat}{3}
    A^q(0) + 4\,\bar{c}^q(0)
    = 2\int d^4k\;
    \biggl(\frac{P\cdot k}{M^2}\, A_2^q 
    & + & \frac{k^2}{M^2}\,&
    A_3^q\biggr)\,,
    \nonumber\\
    A^q(0) + \phantom{4}\,\bar{c}^q(0)
    = 2\int d^4k\;
    \biggl(\frac{P\cdot k}{M^2}\,A_2^q 
    & + & \frac{(P\cdot k)^2}{M^4}\,& 
    A_3^q\biggr)\,.
\end{alignat}
Solving this simple system of linear equations yields
\begin{align}
    A^q(0) 
    = & \frac{2}{3}\int d^4k\biggl(
    \frac{\,2(P\cdot k)^2}{M^4}+\frac{k^2}{M^2}\biggr)\,A_3^q
    + 2\int d^4k\,\frac{P\cdot k}{M^2}\,A_2^q
    \,,
    \nonumber\\
    \bar{c}^q(0)
    = & \frac{2}{3}\int d^4k\biggl(
    - \frac{(P\cdot k)^2}{M^4}+\frac{k^2}{M^2} \biggr)\,A_3^q\,.
\end{align}
So far the expressions are manifestly Lorentz invariant. 
Now we choose to evaluate them in the nucleon rest frame
(\ref{Eq:P-k-rest-frame}) (or in the frame (\ref{Eq:P-k-of-v})
which yields the same result). We obtain
\begin{align}
    A^q(0) 
    = &\int d^4k\biggl(
    2\,\frac{k_0^2}{M^2}+\frac{2}{3}\;\frac{\vec{k}{ }^2}{M^2}\biggr)\,A_3^q
    + 2\int d^4k\,\frac{k^0}{M}\,A_2^q
    \,,
    \nonumber\\
    \bar{c}^q(0)
    = & \int d^4k\biggl(
    - \frac{2}{3}\;\frac{\vec{k}{ }^2}{M^2} \biggr)\,A_3^q\,.
\end{align}
The above expression for $A^q(0)$ was encountered in the momentum sum 
rules in Eqs.~(\ref{Eq:mom2-PDF}-\ref{Eq:mom2-quasi-3}). We also 
recognize that the expression for $\bar{c}^q(0)$ appears in the 
momentum sum rule for the QPDF $D^q(x_v,\gamma^3,v)$ in
Eq.~(\ref{Eq:mom2-quasi-3}). This completes the proof of the momentum
sum rules Eqs.~(\ref{Eq:general-mom1-PDF},~\ref{Eq:general-mom1-quasi-0},~\ref{Eq:general-mom1-quasi-3}).

\end{document}